# Unbiased Inclination Distributions for Objects in the Kuiper Belt

A. A. S. Gulbis[1,2,3], J. L. Elliot[3,4,5], E. R. Adams[3], S. D. Benecchi[6], M. W. Buie[7], D. E. Trilling[8], and L. H. Wasserman[5]

[1]South African Astronomical Observatory, P.O. Box 9, Observatory, 7935 Cape Town, South Africa; amanda@saao.ac.za.
[2]Southern African Large Telescope, P.O. Box 9, Observatory, 7935 Cape Town, South Africa.
[3]Department of Earth, Atmospheric, and Planetary Sciences, Massachusetts Institute of Technology, 77 Massachusetts Ave. Cambridge, MA 02139-4307; jle@mit.edu, era@mit.edu.
[4]Department of Physics, Massachusetts Institute of Technology, 77 Massachusetts Ave. Cambridge, MA 02139-4307.
[5]Lowell Observatory, 1400 W. Mars Hill Rd., Flagstaff, AZ 86001; lhw@lowell.edu.
[6]Planetary Science Institute, 1700 East Fort Lowell, Suite 106, Tucson, AZ 85719; susank@psi.edu.
[7]Dept. of Space Studies, Southwest Research Institute, 1050 Walnut St. #400, Boulder, CO 80302; buie@boulder.swri.edu.
[8] Department of Physics and Astronomy, Northern Arizona University, P.O. Box 6010, Flagstaff, AZ 86011; David.Trilling@nau.edu.

*Short title:* KBO Inclination Distributions

## ABSTRACT

Using data from the Deep Ecliptic Survey (DES), we investigate the inclination distributions of objects in the Kuiper Belt. We present a derivation for observational bias removal and use this procedure to generate unbiased inclination distributions for Kuiper Belt objects (KBOs) of different DES dynamical classes, with respect to the Kuiper Belt Plane. Consistent with previous results, we find that the inclination distribution for all DES KBOs is well fit by the sum of two Gaussians, or a Gaussian plus a generalized Lorentzian, multiplied by sin $i$. Approximately 80% of KBOs are in the high-inclination grouping. We find that Classical object inclinations are well fit by sin $i$ multiplied by the sum of two Gaussians, with roughly even distribution between Gaussians of widths $2.0^{+0.6}_{-0.5}$° and $8.1^{+2.6}_{-2.1}$°. Objects in different resonances exhibit different inclination distributions. The inclinations of Scattered objects are best matched by sin $i$ multiplied by a single Gaussian that is centered at $19.1^{+3.9}_{-3.6}$° with a width of $6.9^{+4.1}_{-2.7}$°. Centaur inclinations peak just below 20°, with one exceptionally high-inclination object near 80°. The currently observed inclination distribution of the Centaurs is not dissimilar to that of the Scattered Extended KBOs and Jupiter-family comets, but is significantly different from the Classical and Resonant KBOs. While the sample sizes of some dynamical classes are still small, these results should begin to serve as a critical diagnostic for models of Solar System evolution.

*Key Words*: comets: general  — Kuiper Belt — minor planets, asteroids  — solar system: general  — surveys

1.  INTRODUCTION

A population of small, icy bodies in the outer Solar System, known as the Kuiper Belt (Edgeworth 1943; Kuiper 1951), is thought to be a relic of our planetesimal disk. Observations of the orbital parameters of Kuiper Belt objects (KBOs) can be used to constrain the dynamical history of the outer Solar System and provide insight into the processes that have shaped the region.

In this work, we derive the current inclination distributions of the KBOs observed by the Deep Ecliptic Survey (DES: Millis *et al*. 2002; Elliot *et al*. 2005).  Inclination distributions have proven to be a particularly good probe for learning more about distant populations. Inclination is the easiest diagnostic orbital element to constrain, and orbital inclinations are likely to be preserved for outer Solar System objects that reach the inner Solar System.

For example, before the observational discovery of a KBO (Jewitt & Luu 1992), the existence of a belt of objects beyond Neptune was hypothesized based specifically on the inclination distribution of short-period comets (Fernandez 1980).  Duncan *et al*. (1988) employed numerical simulations to demonstrate that the inclinations of short-period comets are preserved when these bodies are scattered from the outer Solar System into cometary orbits and thus must stem from a low-inclination population in the outer Solar System.  More recent studies support the Kuiper Belt as a source for short-period comets, specifically proposing that the Jupiter-family comets (JFCs) originated from a scattered disk of KBOs and possibly transitioned as Centaurs (e.g. Holman & Wisdom 1993; Jewitt & Luu 1995; Duncan & Levison 1997; Levison & Duncan 1997; Tiscareno & Malhotra 2003; Emel'yanenko *et al*. 2005; Gomes *et al*. 2008).  Levison *et al*. (2001; 2006) similarly argue that the non-isotropic inclination distribution of the Halley-type comets (HTCs) cannot naturally arise from the Oort cloud – the source region is likely a flattened, scattered disk of objects beyond Neptune. The exact source region(s) for JFCs and HTCs is still under debate. Accurate KBO inclination distributions can therefore serve as important tests of cometary source models.

 KBO inclination distributions are also a critical diagnostic for comprehensive outer Solar System evolutionary models.  Examples include models of Neptune's migration into a quiescent versus pre-excited Belt (Hahn & Malhotra 2005), models involving Neptune "evaders" (Gomes 2003), and the "Nice" model that proposes a rapid migration of the giant planets after a long quiescent period (e.g. Gomes *et al*. 2005; Morbidelli *et al*. 2005; Tsiganis *et al*. 2005; Levison *et al*. 2008).  The results of such dynamical models must be able to reproduce the known KBO inclinations in order to be considered valid descriptions of the evolution of the outer Solar System.

Current surveys of the Kuiper Belt are highly biased because they do not cover the entire sky and/or are constrained by limiting magnitude.  Observational data must be carefully debiased before drawing conclusions or comparing to results from dynamical models. As a controlled set of observations from which to remove observational biases, we use data from the DES. The DES had formal survey status from 2001–2005 by the National Optical Astronomy Observatories (NOAO). The survey was designed to discover and determine the orbits of hundreds of KBOs in order to understand how dynamical phase space is filled in the outermost Solar System. Employing the wide-field Mosaic cameras (Muller *et al*. 1998) on the 4-m Mayall and Blanco Telescopes, the DES imaged over 800 deg$^2$ within ± 6º of the ecliptic. The mean 50% sensitivity of the survey was *VR* magnitude 22.5 (Elliot *et al*. 2005; hereinafter referred to as E05). From its inception in 1998, through the writing of this paper, the DES has discovered nearly 500 designated KBOs. A subsample of these objects has consistent discovery parameters and can be considered for debiasing analyses.

In addition to E05, previous studies of debiased KBO inclination distributions include the groundbreaking work by Brown (2001) and analyses of KBOs discovered at the Canada-France-Hawaii Telescope (Trujillo *et al*. 2001; Kavelaars *et al*. 2008; Kavelaars *et al*. 2009). Here, we use the large, consistent sample of DES KBOs, we explicitly account for survey biases, and we present the first inclination distributions for some dynamical classes.

We begin by introducing a method for removing observational biases due to object orbital inclinations and DES search-frame properties. We employ this method to generate unbiased inclination distributions for DES KBOs (separated by dynamical class) and Centaurs. Next, we investigate functional forms as possible fits to the unbiased inclination distributions. Statistical tests are then applied to compare the unbiased KBO inclination distributions with those of the Centaurs and JFCs. Finally, we discuss the results and compare with previous findings.

## 2. BIAS FACTORS FOR INCLINATION DISTRIBUTIONS

Our method is to calculate the relative likelihood factor for detecting each object discovered on DES search frames throughout the entire survey and to apply this factor to debias the detections. We define likelihood to be a quantity that is proportional to the probability, with the constant of proportion being the same for likelihoods used in the same context. The likelihood of detection is a function of the observational bias due to the geometry of the object's orbit, the position of the search frames, and the properties of the search frames (which are used in pairs). The factors contributing to the likelihood of detection are: (i) the object and search-frame pair geometry, (ii) the solid angle covered by the search-frame pair (accounting for overlap between frames and loss due to interfering objects), and (iii) the limiting magnitude of the search-frame pair. We assume that magnitude and inclination are independent of each other and that objects are on circular orbits. These assumptions are discussed further in §4. Our technique is similar to that employed in §8 of E05; however, the methodology for calculating the geometric, observational bias has been revised (see §2.1).

The biases listed above are based only on the discovery observations. It is worth considering that biases may exist in the recovery observations, which were necessary for orbits to be firmly established to allow designation by the Minor Planet Center (MPC). We address this issue by first looking at the orbital properties of the 369 DES-discovered objects that were not recovered. All but a few of these objects had arc lengths of two days or less, and the orbital parameters were not well established. In fact, we find that only 1% of the lost objects had errors that were smaller than 50% of the nominal values for all three parameters *a, e*, and *i* – 60% of the lost objects had errors > 100% of the nominal value for *i*. Therefore, we cannot say anything conclusive about the orbital parameters of these lost objects. The most likely reasons for nonrecovery are faintness and/or fast motion. Losing faint objects should not effect our debiasing since we consider the magnitude distribution that was derived for the full population (see §2.3). Objects can be fast moving as a result of low *a* or having high *e* and being near perihelion. The DES sample shows no correlation between inclination and semimajor axis (*a* versus *i* for DES KBOs has a Spearman rank-order correlation coefficient of -0.03, following Press *et al*. 2007). The eccentricity of the DES sample is only weakly correlated with inclination (see §4). This implies that losing objects at low *a* or high *e* should not preferentially eliminate objects of any particular inclination. Based on this initial assessment, we conclude that any bias due to lost objects should not have a large impact on our results. A more complete analysis of biases from recovery observations will be carried out in a subsequent DES paper.

Note that in this work we determine the *relative* likelihood of detecting the objects that were discovered by the DES and subsequently designated (i.e. a likelihood of 1 is assigned to the object most likely to have been detected based on the analyses presented here). This method is sufficient for our purposes of debiasing and studying the observed DES inclination distributions. We do not calculate absolute detection probabilities nor account for biases that might have resulted in non-detections. Therefore, we cannot derive unbiased population numbers. We leave this important topic for future work.

*2.1. Geometric component of likelihood*

Consider an object on a circular orbit that is in a chosen reference plane (i.e. ecliptic or invariable), having an orbital pole aligned with the pole of the plane. The probability density over the orbital longitude, *l*, is

$$p(l) = \frac{1}{2\pi}, \qquad 0 \le l \le 2\pi.$$

(1)

To determine the probability of detection over a range of latitudes, *dl*, this value should be integrated as $\int p(l)\, dl$ . The probability density of the orbital longitude is related to the probability density of the orbital latitude, *β*, by

$$p(l)dl = p(\beta)d\beta, \qquad -90° \le \beta \le 90°.$$

(2)

Let the object orbit in the *xy*-plane of an *x, y, z*- coordinate system. We can rotate the system by orbital inclination angle, *i,* with respect to the reference plane (about the *x*-axis), into an *x′, y′, z′*- system:

$$
\begin{aligned}
x &= \cos l & x' &= x \\
y &= \sin l & y' &= y\cos i - z\sin i \\
z &= 0 & z' &= y\sin i + z\cos i
\end{aligned}
$$

(3)

The orbital pole for the inclined orbit is aligned with the *z′*-axis, and as a function of the orbital latitude,

$$z' = \sin\beta.$$
(4)

Combining equations (3) and (4) provides the following relationship between orbital inclination $i$, latitude $\beta$, and longitude $l$:
$$\sin\beta = \sin i \sin l,$$
(5)

which has differential form
$$\cos\beta \, d\beta = \sin i \cos l \, dl.$$
(6)

Employing the Pythagorean identity, equation (5) can be written as
$$\cos l = \sqrt{1 - \frac{\sin^2\beta}{\sin^2 i}}.$$
(7)

By combining equations (1), (2), and (6), the probability density of the orbital latitude can be written as
$$p(\beta) = \frac{\cos\beta}{2\pi \sin i \cos l}.$$
(8)

We combine equations (7) and (8), and normalize such that $\int_{-90°}^{90°} p(\beta'|i) d\beta' = 1$ for any inclination, to define the conditional probability density for finding an object at latitude $\beta$ as a function of its orbital inclination $i$:

$$p(\beta|i) = \begin{cases} \dfrac{\cos\beta}{\pi\sqrt{\sin^2 i - \sin^2\beta}}, & \sin i > |\sin\beta| \\ 0, & \sin i \leq |\sin\beta|, \; i \neq \beta \neq 0 \\ 1, & i = \beta = 0 \end{cases}.$$
(9)

The special cases defined in equation (9) are (i) when $\sin i \leq |\sin \beta|$ and there is zero probability for detecting an object whose inclination never reaches the latitudes of a search-frame pair, and (ii) when both the inclination and the latitude equal zero and the object is detected. Because the probability density is a function of latitude and inclination (and is independent of longitude), the second geometry corresponds to always detecting an object having an orbit of 0° inclination in a DES search-frame pair taken at $\beta = 0°$.

Next, we consider the range of latitudes and longitudes covered by the DES search fields. Every search field on the sky was imaged in at least two Mosaic frames, and each frame was composed of eight CCDs. We separately consider the properties of each CCD on each valid search frame (valid frames are those that overlap in right ascension and declination for each observation in a pair as well as containing all eight Mosaic CCDs). A CCD schematic is displayed in Figure 1. Each CCD ranges from a minimum latitude, $\beta_{min}$, to a maximum latitude, $\beta_{max}$. The tilt of a CCD with respect to the reference plane (at which $\beta = 0°$) is represented by $\theta$. This CCD geometry can be analyzed with respect to different reference planes such as the ecliptic, invariable, or planes determined by objects within the Kuiper Belt.

For each of the CCDs that constitute a DES Mosaic frame, the width, $w$, and height, $h$, equal $0.148°$ and $0.296°$ respectively. These solid angle measurements assume a constant Mosaic plate scale of 0.26"/pixel: the variation across the field is considered negligible, although it decreases quadratically by 6.5% out to the corners. The CCD dimensions can be expressed as $h = \Delta\delta$ and $w = \Delta\alpha\cos\delta_0$, where $\Delta\alpha$ and $\Delta\delta$ are the angular extents of the search field in right ascension and declination and $(\alpha_0, \delta_0)$ is the center of the search field. Employing CCD measurements in terms of angle removes the $\cos\beta$ factor which otherwise would be required to account for change in longitude as a function of latitude.

A convenient CCD measure (cf. Figure 1) is the full latitude extent,
$$H = h\cos\theta + w\sin\theta = |\beta_{max} - \beta_{min}|.$$
(10)

The longitude extent of the CCD, $\Delta\lambda$, as a function of latitude and tilt angle is

$$\Delta\lambda(\beta,\theta) = \begin{cases} 0, & -90° \leq \beta < \beta_{min} \\ \Delta\lambda_{max}(\theta)\dfrac{\beta - \beta_{min}}{\Delta\beta_1(\theta)}, & \beta_{min} \leq \beta < \beta_{min} + \Delta\beta_1(\theta) \\ \Delta\lambda_{max}(\theta), & \beta_{min} + \Delta\beta_1(\theta) \leq \beta < \beta_{max} - \Delta\beta_1(\theta) \\ \Delta\lambda_{max}(\theta)\dfrac{\beta_{max} - \beta}{\Delta\beta_1(\theta)}, & \beta_{max} - \Delta\beta(\theta)_1 \leq \beta < \beta_{max} \\ 0, & \beta_{max} \leq \beta \leq 90° \end{cases}$$

(11)

(following equation B5 of E05), where $\Delta\lambda_{max}(\theta)$ is the maximum longitude within the CCD and $\Delta\beta_1(\theta)$ is a latitude measurement labeled in Figure 1. The measurements used in our analyses, which are functions of the CCD width and height and are labeled in Figure 1, are defined as follows:

$$\Delta\lambda_{max}(\theta) = \begin{cases} \dfrac{h}{\sin\theta}, & \tan\theta \geq h/w \\ \dfrac{w}{\cos\theta}, & \tan\theta < h/w \end{cases},$$
(12)

$$\Delta\beta_1(\theta) = \begin{cases} h\cos\theta, & \tan\theta \geq h/w \\ w\sin\theta, & \tan\theta < h/w \end{cases}, \quad \text{and}$$
(13)

$$\Delta\beta_2(\theta) = \frac{H - 2\Delta\beta_1(\theta)}{2}.$$
(14)

The geometric likelihood component for detecting the *j*th object (which has orbital inclination $0 \leq i_j \leq 180°$) on the *k*th CCD is found by integrating the parameters from equations (9) and (11) over the CCD latitudes:

$$\xi_{\text{lat},k,j} = \int_{\beta_{\min,k}}^{\beta_{\max,k}} \Delta\lambda(\beta',\theta_k) p(\beta' | i_j) d\beta', \tag{15}$$

where $0° \leq \theta_k < 180°$, $-90° \leq \beta_{\min,k} \leq 90°$, and $-90° \leq \beta_{\max,k} \leq 90°$ represent values specific to the *k*th CCD. Equation (15) can be employed for object orbits referenced to various planes, as long as the inclinations and latitudes are self-consistent. Note that evaluation of this integral requires some care – details of our method for solving equation (15) are presented in the online-only Appendix A. Figure 2 provides examples of the resulting geometric likelihood component as a function of selected KBO orbital inclinations, CCD latitudes, and CCD tilt angles.

In this work, we choose to debias using the conditional probability density $p(\beta|i)$, where the inclination is known, rather than $p(i|\beta)$, where the latitude is known (which was the probability function provided in equation 36 of E05). Theoretically, these functions return the same results. The difference is that here we consider the likelihood of detecting an object with known inclination over a range of field latitudes, rather than assuming a latitude and finding the likelihood over a range of inclinations. We select $p(\beta|i)$ in order to be consistent with our sample selection of objects with low inclination errors. In practice, the analytic method that was employed for inclination debiasing in E05 is a close approximation to the more accurate debiasing method used here (see Appendix A online for details).

### 2.2. Solid angle component of likelihood

Although each DES frame covers the same solid angle in the sky, the solid angle over which objects can be detected may be less due to obscuration by other objects and/or misregistration in the centers of the discovery pair of frames. To determine the solid angle component of likelihood, we consider these issues for all CCDs on all valid DES search frames. Defining $\Omega_s$ as the solid angle of a full CCD and $\Omega_k$ as the net solid angle for the *k*th CCD, the solid angle component of the likelihood factor (following E05) is

$$\xi_{\text{ang},k} = \Omega_k / \Omega_s. \tag{16}$$

### 2.3. Magnitude component of likelihood

The magnitude component of the likelihood factor is based on the detection efficiency for each CCD as well as the magnitude distribution of the discovered objects. We describe the detection efficiency using the functional form of Trujillo, Jewitt, & Luu (2001):

$$\varepsilon(m, m_{1/2,k}) = \frac{\varepsilon_{\max}}{2}\left[1 + \tanh\left(\frac{m_{1/2,k} - m}{\sigma_m}\right)\right], \tag{17}$$

where $m$ is the KBO magnitude, $m_{1/2,k}$ is the magnitude on the $k$th CCD at which the detection efficiency has dropped to ½, $\varepsilon_{max}$ is the maximum efficiency at bright magnitudes (here set equal 1), and $\sigma_m$ is a parameter that specifies a characteristic range over which the survey efficiency drops from $\varepsilon_{max}$ to 0. Following E05, we define $m_{1/2,k} = m_{2\sigma,k} - \Delta m_{1/2-2\sigma}$, where $m_{2\sigma,k}$ is the magnitude of an object whose peak pixel is two standard deviations above the mean background for the shallower exposure of a pair of search frames (assuming $m_{2\sigma,k}$ is the same for each CCD on a given frame) and $\Delta m_{1/2-2\sigma}$ is a constant offset that applies to all CCDs in all frames.

In order to debias for individual objects, equation (17) must be considered for each object magnitude and the detection limit in each of the 19056 CCDs in valid DES search frames. Rather, we assume that discovery magnitude and inclination are not correlated (see discussion in §4) and that the observed inclination distribution will be the same regardless of individual object discovery magnitudes. Thus, we need only account for the bias introduced by frame-to-frame variations in limiting magnitude. We approximate the magnitude distribution as a single power law (following Trujillo *et al*. 2001; Bernstein *et al*. 2004, for example) and integrate the detection efficiency over the range of object magnitudes at discovery, $m_{min}$ to $m_{max}$. The resulting magnitude component of the likelihood on the $k$th CCD is

$$\xi_{mag,k} = \alpha \ln 10 \int_{m_{min}}^{m_{max}} 10^{\alpha(m-m_0)} \varepsilon(m, m_{1/2,k}) dm,$$

(18)

(equation 31 of E05) where $\alpha$ is the logarithmic slope of the magnitude distribution and $m_0$ is the magnitude for which the sky density of objects brighter than $m_0$ is one per square degree (at opposition and for $\beta = 0°$). We assume DES parameters $m_{min} = 19.0$, $m_{max} = 25.7$, and the adopted solution for the magnitude distribution from E05: $\sigma_m = 0.58$, $\alpha = 0.86$, $m_0 = 22.70$, and $\Delta m_{1/2-2\sigma} = 1.43$.

### 2.4. Overall likelihood of detection

The likelihood factor for detecting the $j$th object on the $k$th CCD is the combination of the three components described above:

$$\zeta_{detect,k,j} = \xi_{lat,k,j} \xi_{ang,k} \xi_{mag,k}.$$

(19)

The likelihood for detecting the $j$th object over the entire survey is found by summing over $N_F$, the total number of CCDs on all valid DES search frames:

$$\zeta_{detect,j} = \sum_{k=1}^{N_F} \zeta_{detect,k,j}.$$

(20)

### 2.5. Fraction of objects per degree of inclination

Following the bias removal procedures described in E05, we use the likelihood factors from equation (20) to derive the inclination distribution for the fraction of objects per degree, $f_i(i)$. For objects separated into a total of $N_i$ bins, with the $n$th bin having width $\Delta i_n$, we denote the unbiased fraction of the population per degree as $f_{i,n}$. A quantity proportional to the likelihood of detecting an object in the $n$th bin, which contains $N_n$ objects, is $Z_n$:

$$\frac{1}{Z_n} = \frac{1}{N_n} \sum_{j=1}^{N_n} \frac{1}{\zeta_{\text{detect},j}}.$$

(21)

The number of objects detected in the $n$th bin can be written in terms the total number of objects in the Kuiper Belt, $N_T$, the unbiased fraction in the bin, the bin size, and a constant $\gamma$ that is proportional to the likelihood of detection:

$$N_n = N_T f_{i,n} \Delta i_n Z_n \gamma,$$

(22)

where $\gamma$ is normalized over all $N_i$ bins:

$$\gamma = \frac{1}{N_T} \sum_{n=1}^{N_i} \frac{N_n}{Z_n}.$$

(23)

Combining equations (22) and (23) provides the following expression for the unbiased fraction of objects in the $n$th bin:

$$f_{i,n} = \frac{N_n}{\Delta i_n Z_n \sum_{n=1}^{N_i} \frac{N_n}{Z_n}}.$$

(24)

The error on the fractional number of objects in each bin, $\sigma(f_{i,n})$, is estimated as

$$\frac{\sigma(f_{i,n})}{f_{i,n}} = \frac{\sigma(N_n)}{N_n}.$$

(25)

Assuming the standard deviation in the number of detected objects per bin follows binomial statistics, with the number of trials being the total number of objects that could have been detected in each bin, $N_T f_{i,n} \Delta i_n$, and a probability of success, $Z_n \gamma$, the error is

$$\sigma(f_{i,n}) = \frac{f_{i,n} \sqrt{N_T f_{i,n} \Delta i_n Z_n \gamma (1 - Z_n \gamma)}}{N_n} \approx \frac{f_{i,n}}{\sqrt{N_n}}.$$

(26)

Without knowing the total number of objects in the Kuiper Belt we cannot accurately evaluate $\gamma$; however, $Z_n \gamma$ is expected to be $\ll 1$ assuming tens of thousands of total objects.

Note that the fractional error given in E05 equation (9) incorrectly considers the square root of the debiased number rather than the detected number of objects in the $n$th bin. The difference between the fractional errors calculated using equation (26) and E05 equation (9) is directly proportional to the number of objects per degree in each bin. As a result, the error bars on the data presented in E05 are too small for bins that have less than approximately 10 objects per degree of inclination and are too large for bins containing more objects. This corresponds to the error bars in E05 figures 16, 18, and 20a being 40% to 80% too small on binned data points with $i_K > 8°$ and being 105% to 160% too large on binned data points with $i_K < 8°$.

## 3. RESULTS

### 3.1. Unbiased inclination distributions

The database of DES objects (Buie *et al*. 2003) contained 482 KBOs and Centaurs on valid search fields as of 2007 October 24. The survey has officially ended, and this sample has not changed significantly between that time and submission of this paper. From this sample, we select only the objects with low error on their orbital inclination (≤ 0.5°) to allow for higher accuracy in the derived inclination distributions. We also consider subsamples of this population based on dynamical classification. The samples we consider, and the number of objects in each sample, are listed in Table 1. We note that there are no retrograde objects in our samples. This is not due to the exclusion of such objects in our recovery efforts, or a bias in our discovery fields: no retrograde objects were detected.

We follow the DES classification scheme presented in E05, which tests for the following criteria in the order listed: Resonant objects are those for which one or more resonant arguments librate; Centaurs are objects whose osculating perihelia reach values less than the osculating semimajor axis of Neptune; Scattered Near objects have time-averaged Tisserand parameters relative to Neptune of $T_N < 3$; Scattered Extended objects have $T_N > 3$ and time-averaged eccentricities > 0.2; and Classical objects have $T_N > 3$ and time-averaged eccentricities < 0.2. Objects are classified by assuming three sets of initial conditions that describe a nominal orbit, along with two clones at ± one sigma excursions in semimajor axis and eccentricity space, and integrating for 10 Myr (see E05 for a complete description). If all three initial-condition sets do not return a classification because the errors in the orbital elements are too large, the object is deemed Unclassified. Some Unclassified objects have low inclination errors and thus meet the criteria to be considered in our analyses. We stress that the Unclassified objects are not the same as those that are lost: this sample has well-defined inclinations. As noted by Gladman *et al*. (2008b), inclinations are tied to the Tisserand parameter (i.e. higher inclinations return lower $T_N$). The DES Scattered Near and Scattered Extended classifications thus preferentially sort objects by inclination. To overcome this partiality, we also consider the grouping of the combined Scattered objects in our analyses.

A natural reference for KBO orbits is the mean orbital plane of the Kuiper Belt itself. Here we reference inclinations to the Kuiper Belt plane (KBP) for debiasing, as parameterized by the adopted solution in E05 (see Appendix B for conversion of object orbital inclination between planes). With respect to the plane of the ecliptic, the KBP has inclination 1.74° ± 0.23° and node 99.2° ± 6.6°, which is consistent with the invariable plane of the Solar System. We represent orbital inclinations with respect to the KBP as $i_K$. The maximum difference between $i_K$ and ecliptic inclination in our sample of DES KBOs and Centaurs is 1.62°. Note that slightly different locations for the KBP have been calculated (theoretically and observationally: Brown & Pan 2004; Chiang & Choi 2008). In fact, it is expected that each dynamical class has its own reference plane (E. Chiang, personal communication). The framework described here can be applied to any reference plane.

As an illustration of the bias factors, the normalized detection likelihoods for the DES KBOs are plotted in Figure 3. The dominance of the geometric likelihood component, $\xi_{lat}$, is apparent as a general decrease in detection likelihood with increasing KBP inclination. Below approximately 0.5º, the likelihood of detection is relatively lower due to KBO orbits not spending as much time in the latitudes of the discovery search fields.

The observed inclination distribution for the 344 DES KBOs that have low inclination errors is shown in Figure 4*a*. The data are plotted per degree: the sum of the data points in each bin multiplied by the bin widths equals the total (1 for the fraction of objects labeled on the left and 344 for the number of objects labeled on the right). These observations are highly biased. From the debaising method in §2, the unbiased inclination distribution for the full KBO sample is shown in Figure 4*b*. The strength of the bias against detecting high-inclination objects is apparent in the difference between Figures 4*a* and 4*b*. To provide a visual representation of the types of KBOs that constitute this distribution, the bins in Figure 4 are shaded to reflect the proportion of objects by dynamical class. Figure 4b is not labeled by number of objects, since our debiasing only accounts for objects discovered by the DES and does not attempt to derive total population numbers.

The double-peaked nature of the total inclination distribution for KBOs that was originally reported by Brown (2001), and confirmed by E05, is apparent. In comparison to this work, the unbiased inclination distribution presented in E05 considered a similarly derived sample of 240 DES objects (see Table 1 for additional sample comparisons).

The unbiased inclination distribution for the 17 DES Centaurs having low inclination errors is displayed in Figure 5. We consider the Centaurs separately from KBOs due to the perturbed nature of their orbits: as "planet-crossers," Centaurs are expected to have dynamically short lifetimes. The small number of Centaurs in this sample does not yet allow for a well-defined inclination distribution. However, Figure 5 suggests that Centaur inclinations peak at less than 20º. This peak is at a slightly lower inclination, and is significantly less pronounced, than that found by Emel'yanenko *et al*. (2005). There is only one Centaur composing the highest inclination bin, and it is the DES object with the highest inclination ($i_K$ = 76.5º; ecliptic inclination 78.0º).

There are now enough classified DES KBOs that it is worthwhile to investigate the inclination distributions for different dynamical classes. Binned inclination distributions for the considered samples are displayed on the left side of Figure 6. On the right, cumulative distributions are shown in order to provide a representation of unbinned data.

As noted by Brown (2001), KBOs of different dynamical classes have strikingly different inclination distributions. The Classical objects have a double-peaked distribution, and they dominate the low-inclination grouping in the distribution for DES KBOs. In general, Resonant objects have a fairly flat inclination distribution extending to ~ 30º. The differences in the distributions of the two most-populated resonances (3:2 and 7:4) demonstrate that inclinations vary by resonance. Scattered objects dominate the higher-inclination portion of the distribution for DES KBOs. Scattered Near objects have inclinations greater than 10º with a distribution peak just above 20º and Scattered Extended objects have inclinations less than 20º. These distributions demonstrate the tie between Tisserand parameter and inclination, and thus the selection effect of our classification scheme. The Unclassified inclination distribution contains primarily low-inclination objects and is most similar to the Classicals – this is not surprising, given that the likelihood of detection is higher for low-inclination objects and that more than half of the classified objects in our sample are Classicals.

### *3.2. Fitting method*

To investigate functional forms for the inclination distributions, we perform Monte Carlo simulations and use Kuiper's variant of the Kolmogorov-Smirnov (K-S) test. Kuiper's variant employs the maximum deviations (plus and minus) between two cumulative distributions to quantify the probability that a random sample would result in a larger difference between the two samples than that observed. We select Kuiper's test because it is effective at detecting changes in the tails of the distributions, whereas the standard K-S test is more sensitive to shifts in the distributions (Press *et al*. 2007). The statistic for this test is $V\sqrt{N}$, where $V$ is Kuiper's statistic and $N$ is the effective number of objects in the compared datasets as defined in Press *et al*. (2007). Following Brown (2001), we calibrate $V\sqrt{N}$ by comparing a uniform distribution to 100,000 datasets comprised of random numbers distributed between 0 and 1. The probability of obtaining a particular value of $V\sqrt{N}$ is determined using the cumulative probability, as shown in Figure 4 of Brown (2001). The 1-, 2- and 3-sigma levels correspond to cumulative probabilities of 84.1%, 97.7%, and 99.9%, and $V\sqrt{N}$ values of 1.49, 1.85, and 2.29 respectively.

For our Monte Carlo method, we create a distribution of objects following a functional form. We then generate a few thousand simulated datasets from the distribution and employ a K-S test to compare the simulated data to the sample. We implement the K-S test by comparing piecewise continuous functions. The average $V\sqrt{N}$ for all simulations is taken to be the statistic for that functional form. By stepping through a range of values for each of the parameters in the functions, we arrive at a best fit (minimum $V\sqrt{N}$). The best-fit $V\sqrt{N}$ statistic is then compared with the calibrated values to determine the confidence level at which we can reject the null hypothesis, namely that the simulated datasets and the sample were drawn from the same parent distribution. Results at confidence levels less than 1-sigma ($V\sqrt{N}$ < 1.49) correspond to < 84.1% probability that $V\sqrt{N}$ would be larger than that obtained. For all samples, we find functional forms that can only be rejected at confidence levels less than 1-sigma. We thus consider these fits acceptable.

### *3.3. Functional forms for KBO distributions*

Various analytical functions can be considered in trying to match the observed inclination distributions. As demonstrated by Brown (2001), a Gaussian appears to be a natural functional form for the ecliptic inclination distribution of KBOs (multiplied by sin *i* for the total inclination distribution). The distribution of known asteroids is well fit by this form, as are the results from simulations of multiple perturbations in an initially zero-inclination disk (Brown 2001).

Because some of our unbiased inclination distributions fall off rather steeply at high inclinations, we follow E05 and also investigate a generalized Lorentzian multiplied by sin *i*. We do not have any underlying physical reason for this form. For distributions that are clearly double-peaked, we assume that there are overlapping groups of objects and thus consider the sum of two Gaussians or a Gaussian plus a Lorentzian.

We define the variables $G$ and $L$ to represent the following Gaussian and generalized Lorentzian functions:

$$G(i,\sigma,\mu) = e^{-\frac{(i-\mu)^2}{2\sigma^2}} \sin i$$

$$L(i,I,g) = \left[1 + (i/2I)^g\right]^{-1} \sin i \quad (27)$$

Normalized versions of these functions, $G_{norm}$ and $L_{norm}$, are obtained by dividing by the integral of each function over all inclinations:

$$G_{norm}(i,\sigma,\mu) = \frac{G(i,\sigma,\mu)}{\int_{i'=0°}^{180°} G(i',\sigma,\mu) di'}$$

$$L_{norm}(i,I,g) = \frac{L(i,I,g)}{\int_{i'=0°}^{180°} L(i',I,g) di'} \quad (28)$$

Following previous work, we first consider $\sin i$ multiplied by a single Gaussian of width $\sigma_1$ and centered at $\mu_1$,

$$f_G(i) = G_{norm}(i,\sigma_1,\mu_1). \quad (29)$$

The most simple case is where the Gaussian is centered on the plane, $\mu_1 = 0$. We employ our Monte Carlo method and find the range of $\sigma_1$ for which $V\sqrt{N}$ of the simulated datasets relative to the sample is less than the 1-sigma level. For the case of equation (29) with $\mu_1 = 0$, acceptable fits are found for samples of 3:2, 7:4, Other Resonant, and Scattered Extended. Note that the Other Resonant sample consists of all Resonant objects except for those in the 3:2 resonance (Table 1). Best-fit values and 1-sigma error bars are listed in Table 2. The model fitting results are plotted in Figure 7, where best fits and rejection levels are clearly denoted. This functional form can be rejected for all other samples at confidence levels listed in Table 3.

For functions containing two or more parameters, such as equation (29) with a Gaussian offset from the plane ($\mu_1 \neq 0$), the best fit is a multi-dimensional space enclosed by the contour where $V\sqrt{N}$ is at the 1-sigma level. Because the parameters are correlated, we cannot simply take the 1-sigma values as the error bars (see description in Brown 2001). We use our Monte Carlo method and generate contour plots of $V\sqrt{N}$ to determine the nominal best fit and then calculate 1-sigma errors on each parameter while keeping the other parameter(s) fixed.

As listed in Table 2, acceptable fits for equation (29) with $\mu_1 \neq 0$ are found for the samples of 3:2, 7:4, Other Resonant, Scattered, Scattered Near, and Scattered Extended. The fitting contour plots are shown in Figure 8. Although the best-fit value for each sample is marked by a dot, the regions for which the fits are acceptable lie within the 1-sigma contour lines. The samples that have acceptable fits for $\mu_1 = 0$, and the corresponding range of acceptable $\sigma_1$ values, are apparent here. Note that the samples of Scattered and Scattered Near are similar, since Scattered Near comprises the majority of the total Scattered sample.

A function that decreases more steeply at higher inclinations is $\sin i$ multiplied by a single generalized Lorentzian, where the Lorentzian half-inclination is $I_1$ and power is $g_1$:

$$f_L(i) = L_{norm}(i, I_1, g_1).$$
(30)

We find acceptable fits for this functional form to the samples of 3:2, 7:4, Other Resonant, Scattered, Scattered Near, and Scattered Extended. For 7:4 and all three Scattered samples, this form has a wide range of acceptable $g$-parameter space – once the drop-off at higher inclinations becomes steep, higher values of $g$ do not alter the Lorentzian distribution significantly and $V\sqrt{N}$ does not change. An example is provided in Figure 9, for equation (30) fit to the Scattered KBOs.

As listed in Table 2, the fits to 7:4 and Other Resonant have a lower confidence level of rejection for equation (30) than equation (29). The remaining samples have a higher confidence level of rejection for this functional form. While these confidence levels indicate a higher or lower likelihood of the samples being drawn from one functional form versus the other, none of the best fits listed in Table 2 can be ruled out at the 1-sigma level.

For the double-peaked distributions, we consider a function that is $\sin i$ multiplied by the sum of two Gaussians centered on the plane,

$$f_{2G}(i) = a_1 G_{norm}(i, \sigma_2, 0) + (1 - a_1) G_{norm}(i, \sigma_3, 0),$$
(31)

where the widths are represented by $\sigma_2$ and $\sigma_3$, and $a_1$ is the fraction of objects in the first Gaussian. Acceptable fits are found for DES KBOs, Classical, and Unclassified samples. As an example of our method for fitting a three-parameter functional form, Figure 10 contains a series of contour plots showing the confidence levels as a function of $\sigma_2$ and $\sigma_3$ for different values of $a_1$. This example is for the sample of Classical KBOs.

Finally, we consider $\sin i$ multiplied by a Gaussian plus a generalized Lorentzian, where $a_2$ is the fraction of objects in a Gaussian of width $\sigma_4$, and the Lorentzian half-inclination is $I_2$ and power is $g_2$:

$$f_{GL}(i) = a_2 G_{norm}(i, \sigma_4, 0) + (1 - a_2) L_{norm}(i, I_2, g_2).$$
(32)

Acceptable are found for DES KBOs, Classical, and Unclassified samples. Figure 11 contains two contour plots showing the confidence levels surrounding the best-fit result for the Classical KBOs. The best-fit functional forms listed in Table 2 are plotted along with the unbiased inclination distributions in Figure 6.

### 3.4. Source region comparison

The idea that the Kuiper Belt may serve as a source region for JFCs, with the Centaurs as a transition population, can be investigated by comparing inclination distributions. As a first-order evaluation, we compare the debiased DES Centaur inclination distribution with those of different samples using Kuiper's variant of the K-S test as described in §3.2. To calibrate $V\sqrt{N}$ for comparing two samples of known sizes, we generate a uniform distribution between 0 and 1 and draw two simulated datasets from it containing the same number of objects as each of the compared datasets. We repeat this process tens of thousands of times to calculate the probability of obtaining a particular value of $V\sqrt{N}$ and thus the sigma confidence levels (the same procedure as described for a uniform distribution in §3.2).

The cumulative distribution of Centaur inclinations (solid data points in Figure 5b) is compared with those of the KBO dynamical samples (shown on the right side of Figure 6) and that of the JFCs (open data points in Figure 5b). The JFC sample contains 256 MPC comets that have Tisserand parameters with respect to Jupiter such that $2 < T_J < 3.05$ (following Gladman *et al*. 2008b) and perihelia $q < 2.5$ AU (following Levison & Duncan 1997). The inclination distribution of these comets is considered to be unbiased because they approach the Earth closely enough to have been well observed, at many different locations, over a long period of time.

Results from our K-S tests are listed in Table 4. We find that the null hypothesis, that the currently observed Centaur inclinations were drawn from the same parent distribution as the samples of DES KBOs and Scattered Extended, can be rejected at a confidence level less than 1-sigma. Therefore, it is statistically possible that these samples have the same intrinsic inclination distribution. For comparisons between Centaurs and JFCs, the null hypothesis is rejected at a confidence level between 1 and 2 sigma. For comparisons between Centaurs and the remaining samples, the null hypothesis is rejected at confidence levels > 2 sigma.

The JFC inclinations are also most closely matched by Scattered Extended KBOs, with the probability that they are drawn from the same parent distribution being rejected at a confidence level of between 1 and 2 sigma. Comparisons between JFCs and the remaining samples can be rejected at levels of > 2 sigma.

## 4. DISCUSSION

The debiasing procedure presented in §2 is critical to our investigation, and it contains the assumption of circular orbits. Since KBO and Centaur orbits are elliptical, some being highly eccentric, we must consider how this assumption will affect our results. The eccentricity of our sample of 482 DES KBOs is only weakly correlated with inclination (having a Spearman rank-order coefficient of 0.30, at significance level > 99.9%; Press *et al*. 2007). Thus any bias in eccentricity should fairly evenly affect the entire range of inclinations in our full sample. The variation from the circular assumption would be most pronounced for objects in high eccentricity–high inclination orbits, because these objects have the greatest difference in time spent at different inclinations. The majority of objects do not have both high eccentricity and high inclination (only 13% have both $e > 0.2$ and $i_K > 10°$).

One way to quantify an eccentricity bias would be to simulate objects with highly eccentric orbits and determine the likelihood of detection in all of the DES search fields. That is beyond the scope of this work. We instead consider the Monte Carlo modeling carried out by Brown (2001), which demonstrated that their method derives reasonably accurate inclination distributions even for objects with high eccentricity. In Brown (2001), the key to determining inclinations is the object's orbital inclination and the latitude at discovery. We use a similar assumption of circular orbits to debias for these parameters (compare our equation 9 to equation 3 from Brown (2001)). The robustness of the fitting technique is identical, since we follow Brown (2001) by employing Monte Carlo simulations and K-S testing for our fitting methodology. The similarity of these key analysis components suggests that the general result from Brown (2001) is applicable here. We thus conclude that effects due to orbital eccentricity should not greatly skew our results. However, we note that dynamical samples based on classifications that selectively consider objects of high eccentricity (such as Scattered Extended, Resonant, and Centaurs) would be most influenced by our assumption of circular orbits.

We assume that discovery magnitude and inclination are not correlated, such that a detection efficiency preferential to brighter objects will not affect our derived inclination distributions. While Levison and Stern (2001) determined that there is a correlation between KBO inclination and absolute magnitude, any bias in this work will be associated with discovery magnitude. A Spearman rank-order test (Press *et al.* 2007) between inclination and discovery magnitude supports our assumption: we find a correlation coefficient of 0.01 for classified DES KBOs and correlation coefficients of absolute value < 0.25 for all samples we consider except Centaurs (which have a coefficient of –0.33 at a 25% significance level) and Scattered Extended KBOs (which have a coefficient of 0.71 at a 99% significance level). There are only nine Scattered Extended KBOs, so this correlation may be the result of having only a small number of objects. If the correlation were real, then we may have underestimated the fraction of higher inclination objects in this sample.

We stress that this work derives relative likelihoods of detection. The derivation of the total numbers of objects, and comparison of class populations, requires careful consideration of the orbital eccentricity of each object as well as the detection probability for each object based on its magnitude. In addition, bins containing zero objects here indicate that no DES objects were discovered, *not* that no objects exist at those inclinations. These analyses are reserved for future work.

*4.1 Comparison with previous findings*

We have found statistically good fits for the functions presented in see §3.3 to the unbiased inclination distributions for all considered KBO samples. The distributions for the samples of DES KBOs, Classicals, and Unclassified are clearly double peaked while the others are single peaked. Specific samples are discussed in the following subsections.

*4.1.1. DES KBOs*

For the inclination distribution of DES KBOs, we find acceptable fits to sin *i* multiplied by either the sum of two Gaussians or a Gaussian plus a Lorentzian. The double-Gaussian fit is good, being rejected at a confidence level of only 67.9%. The other fit can be rejected at a confidence level of 43% and is thus statistically more likely than the double Gaussian, although neither fit can be ruled out. These results suggest that there may be a steeper drop-off of objects at higher inclinations than expected from a Gaussian distribution. While we do not propose any physical basis for the functional form of a Gaussian plus a Lorentzian, it does provide the best match to the data that we have found.

To compare with previous results, we consider our best-fit double Gaussian. The results are consistent with E05 to within the error bars. Compared to Brown (2001), which determined Gaussians of widths of $2.6^{+0.8}_{-0.2}$° and 15° ± 1°, our results indicate slightly smaller widths for both components. The variation between inclinations referenced to the ecliptic versus the mean plane of the Kuiper Belt can result in this disparity of a few degrees. Given the differences in samples and methods between this work and Brown (2001), the overall results are similar and a double-peaked form for the inclinations of all KBOs seems assured.

*4.1.2. Classical*

We find roughly comparable fits to the Classical objects for functional forms of sin *i* multiplied by either the sum of two Gaussians or a Gaussian plus a Lorentzian. The Gaussian plus Lorentzian again has a slightly lower confidence level of rejection than the double Gaussian. For the sake of comparison with previous results, we consider our best-fit double Gaussian. This result is consistent with the Classical low-inclination Gaussian widths of $2.2^{+0.2}_{-0.6}$° and 1.5° ± 0.4° determined respectively by Brown (2001) and Kavelaars *et al*. (2008). However, our high-inclination Gaussian width is less than the values of 17° ± 3° determined by Brown (2001), 13° ± 3° from Kavelaars *et al*. (2008), and ~ 15° from Kavelaars *et al*. (2009). We also find that Classical objects are roughly evenly distributed between the high- and low-inclination groupings. This distribution is significantly different from the majority of objects being in the high-inclination grouping (81% as determined by Brown 2001), but is within the 1-sigma contour of the recent model fits from Kavelaars *et al*. (2009; whose model assumed a low-inclination Gaussian of width 2.2° following Brown (2001)).

The cause for these discrepancies could be partially due to different classification schemes. For example, some objects considered "classical" by Brown (2001) due to low semimajor axes and eccentricities would be classified by the DES as Scattered Near. We also limit Classical objects to DES-discovered KBOs having eccentricity < 0.2. Most likely, our view of the inclination distribution is evolving due to the discovery of many new KBOs since Brown (2001).

*4.1.3. Resonant*

The unimodal functional forms of sin *i* multiplied by either a single Gaussian or by a generalized Lorentzian provide acceptable fits to the inclination distributions of the subsamples of Resonant objects. This general form is consistent with distributions for objects in the 3:2 resonance, Plutinos, from Brown (2001) and Kavelaars *et al*. (2008). Our best fit for the Plutinos is for a single Gaussian centered off the plane. Fits of comparable confidence level are found for a Lorentzian or a single Gaussian centered on the plane. The last result is consistent with the best fit for sin *i* multiplied by a single Gaussian of width $10^{+3}_{-2}$° from Brown (2001) and disagrees with the width of ~15° from Kavelaars *et al*. (2009). The result from Kavelaars *et al*. (2009) is weighted more strongly toward higher inclinations, since three of their eight Plutinos have *i* > 27° whereas all of our 51 Plutinos have *i* < 23°. This difference may indicate that the sample sizes of controlled KBO surveys to date are not yet large enough to fully account for the inclination distributions of these objects.

For objects in the 7:4 resonance, the best fit is a generalized Lorentzian; however, very good fits (rejection confidence level < 20%) are also found for a single Gaussian centered either on or off the plane. The 7:4 distribution is striking in contrast to the Plutinos since all objects in the 7:4 resonance have inclinations < 6º. We generated $10^6$ simulated datasets from the best-fit functional forms to the Plutinos, and zero of them had all objects with inclinations < 6º. Thus the probability that KBOs in the 7:4 resonance have the same inclination distribution as the Plutinos is negligibly low. We conclude that different resonances can exhibit uniquely dissimilar inclination distributions. This is intriguing, given that there are also indications that KBOs in different resonances exhibit different colors (Gulbis *et al*. 2006).

For the sample of Other Resonant objects, we find the best fit for a generalized Lorentzian. An alternatively good fit is found for a single Gaussian centered on the plane. The disparity between inclination distributions for objects in the 3:2 and 7:4 resonances suggests that each resonance should be considered individually rather than being combined into a single sample.

All three of the considered Resonant subsamples have acceptable fits to a single Gaussian multiplied by sin *i* centered on the plane. Since this is a natural physical form, it could indeed represent the true distributions. The fact that we have found statistically more likely fits with the forms of Lorentzians and offset Gaussians may point toward specific shaping of these samples due to resonance sweeping or other events in their dynamical history.

### *4.1.4. Scattered*

The DES samples of Scattered Near, Scattered Extended, and the combined grouping of Scattered have single-peaked inclination distributions. This general form is consistent with distributions for "scattered" from Brown (2001) and for "Scattered" and "Detached" from Kavelaars *et al*. (2008; based on the classification scheme by Gladman *et al*. 2008).

The Scattered Near sample has a distinctive distribution consisting of all high-inclination objects (*i* > 5º). This is an artificially "hot" grouping as a result of classification according to low Tisserand parameter (and thus interaction with Neptune). Comparably good fits are found for sin *i* multiplied by a single Gaussian centered off the plane, or a single Lorentzian.

The Scattered Extended objects are best fit by sin *i* multiplied by a single Gaussian. Acceptable fits are also found for a single Lorentzian and a single Gaussian centered on the plane. The Scattered Extended sample is comparable to the "scattered" groupings from (i) Brown (2001), who found a poor-quality single-Gaussian fit of width 20º ± 4º, and (ii) Kavelaars *et al*. (2008), who reported a double-Gaussian fit in which the low-inclination Gaussian could be non-existent (width 1.6º ± 1.6º) while the high-inclination Gaussian had width 13º ± 5º.

When we consider a combined Scattered sample, the best functional fit is for sin *i* multiplied by a single Gaussian having width 6.9º and centered on 19.1º. An acceptable fit is found for a single Lorentzian. Since the Scattered sample consists of nearly 80% Scattered Near objects, the distributions are quite similar, with Scattered being slightly wider and shifted toward lower inclination. Our Scattered and Scattered Near samples are distinct from the other single-peaked samples in that sin *i* multiplied by a single

Gaussian centered on the plane does not provide an acceptable fit. Reconsideration of the classification scheme is merited, including investigation of "Scattered," "Scattering," or "Detached" as proposed by Gladman *et al.* (2008).

In §4.1.1-§4.1.4 we discuss how different functional forms provide good fits to the considered samples. The acceptability of single Gaussians centered off the plane supports the idea mentioned in §3.1 that dynamical classes may have different reference planes. This is an avenue for future investigation. Larger sample sizes, as we expect from the upcoming Pan-STARRS (Panoramic Survey Telescope and Rapid Response System) and LSST (Large Synoptic Survey Telescope), will also allow more stringent constraints and discrimination between functional forms.

*4.1.5. Centaurs*

In addition to KBOs, there are 17 Centaurs in our sample spread over a wide range of inclinations. Our distribution peak near 20° is consistent with Emel'yanenko *et al.* (2005). However, our Centaur inclinations are more evenly distributed amongst values less than 20° and our distribution extends to 40°. Centaurs at higher inclinations, out to 40°, are supported by numerical models of Tiscareno & Malhotra (2003), but our statistics are still too sparse for definitive results at these values. Similarly, the large error bar on detecting the sole DES object at very high inclination (~80°) makes it difficult to discern the fraction of objects that would be expected in this region. There is a significant observational bias against detecting high-inclination objects, and there are currently only two other known Centaurs with inclination > 45°: 2008 $YB_3$ (McNaught *et al.* 2008) and 2008 $KV_{42}$ (Gladman *et al.* 2008a). It is possible that these high-inclination objects are part of a class that should be considered separately from the Centaurs.

*4.2 Relationships between KBO, Centaur, and Comet inclinations*

Direct comparison between current unbiased inclination distributions using Kuiper's variation of the K-S test suggests connections between KBOs, Centaurs, and JFCs. The evolutions from Scattered objects to JFCs (e.g. Duncan & Levison 1997) and Centaurs to JFCs (e.g. Tiscareno & Malhotra, 2003) have been established by numerical simulations. Transitions have also been demonstrated for JFCs back to Centaurs (Hahn & Bailey 1990; Horner *et al.* 2004) and for limited excursions of Centaurs to Scattered KBOs (Tiscareno & Malhotra 2003). Our findings of the similarities between currently observed JFC, Centaur, and Scattered Extended inclination distributions support these connections. In particular, the possible similarity between Scattered Extended and the Centaur and JFC inclination distributions is consistent with the scattered disk as a specific source region for these groups (following, for example, Duncan & Levison 1997; Levison & Duncan 1997; Duncan *et al.* 2004). Classical and Resonant KBOs have also been suggested as source regions for the JFCs (most recently by Volk & Malhotra 2008). Our results indicate that these KBOs currently have very dissimilar inclination distributions to both the Centaurs and JFCs.

Note that the direct comparison between inclinations of KBOs, Centaurs, and JFCs is based on the assumption that the orbital inclinations remain essentially unperturbed during evolution between the groupings. Studies have shown that the inclination distribution of visible comets roughly follows that of the source population (Duncan *et al*. 1988; Levison *et al*. 2001). However, some numerical simulations have shown evolution in the orbital parameters of Centaurs and comets: Centaur inclinations can change by at least a few degrees over their lifetimes (Tiscareno & Malhotra 2003), and JFC inclinations may be independent of precursor KBO orbits (Levison & Duncan 1997). Therefore, such comparisons serve only as a first-order diagnostic to indicate similarities between currently observed inclinations of these samples.

Many groups (e.g. Peixinho *et al*. 2003; Tegler *et al*. 2003; Peixinho *et al*. 2004; Doressoundiram *et al*. 2005; Delsanti 2006) have demonstrated a bimodal color distribution for Centaurs. This dichotomy could be attributed to multiple source regions or evolutionary paths (Tegler *et al*. 2008). Indeed, of the two groups of Centaurs identified by Lamy and Toth (2009), Centaurs I (red) have significantly smaller inclinations than Centaurs II (gray). While there are no obvious connections between the colors of specific KBO dynamical classes and those of the two Centaur groupings (Lamy & Toth 2009), and this work reveals only a possible similarity between the current inclination distributions of Scattered Extended KBOs (known to have a range of colors; Gulbis *et al*. 2006) and all Centaurs combined, future analyses of inclinations and colors could prove productive in deriving Centaur source regions.

### *4.3 Implications for evolutionary models*

The inclination distribution of KBOs records and reflects the early evolution of the outer Solar System. Several models of the history of the outer Solar System have been employed in an attempt to understand the observed properties of the Kuiper Belt. Hahn and Malhotra (2005) were able to reproduce the population of low-inclination KBOs quite successfully by using a model in which Neptune migrates outward into a sea of proto-Kuiper Belt particles. However, this model failed to reproduce the high-inclination population. In comparison, Levison *et al*. (2008) uses the "Nice" model, in which encounters among the outer planets emplace objects into the Kuiper Belt gravitationally. This model does relatively well in reproducing both the low- and high-inclination KBO populations (see their Figures 6, 7, and related figures), though in detail their model results are not in excellent agreement with the observed inclination distribution (see their Figure 12 as well as our Figure 6, topmost panel on the right).

At present, there is not a clear, general theoretical model that reproduces the overall KBO inclination distribution. The inclination distributions that we produce for the various subpopulations (Figure 6) are likewise not matched by any current theoretical model. These distributions offer stringent constraints on the history of the outer Solar System and likely record specific timings and origins of these populations. We look forward to the next generation of theoretical models that can use our entire suite of results as their comparative ground truth.

## 5. CONCLUSIONS

This work presents an analysis of KBO inclination distributions based on data from the DES. Our sample contains 344 DES KBOs and 17 Centaurs having errors in inclination of $\leq 0.5°$, and all inclinations are referenced to the mean plane of the Kuiper Belt as derived in E05. Our conclusions are enumerated below.

1. Debiasing procedures are essential in analyzing survey data. We present a detailed method for debiasing DES data in order to specifically study KBO inclinations. This method is generalized such that inclinations can be referenced to a chosen plane.
2. A double-peaked distribution for DES KBOs is confirmed, with approximately 80% of the objects in the higher-inclination grouping. We also confirm a double-peaked distribution for Classical KBOs, which is well fit by sin $i$ multiplied by the sum of two Gaussians with roughly an equivalent fraction of objects in each Gaussian. The functional form of sin $i$ multiplied by a Gaussian plus a generalized Lorentzian provides fits that are statistically more likely for these samples, possibly indicating a steeper drop than expected at higher inclinations.
3. Different DES dynamical classes exhibit distinct inclination distributions. Objects in resonances should be considered separately, as demonstrated by remarkably different distributions for objects in the most populated DES resonances of 3:2 and 7:4. Scattered Near, Extended, and the combined Scattered sample are well fit by sin $i$ multiplied by either a single Gaussian offset from the plane (by ~17°-20°) or a generalized Lorentzian. Scattered Near and Scattered samples are the only single-peaked groupings considered for which no acceptable fits were found for the functional form of sin $i$ multiplied by a single Gaussian centered on the plane.
4. A simple statistical comparison between current inclination distributions shows that, at a rejection level of < 2 sigma, the parent distribution could be the same for (i) Centaurs and Scattered Extended and DES KBOs, (ii) Centaurs and JFCs, and (iii) JFCs and Scattered Extended KBOs. The current inclination distributions of Resonant and Classical KBOs are statistically dissimilar to the JFCs and Centaurs.

Although sample sizes remain low for some dynamical classes, the unbiased inclination distributions presented here should serve as useful diagnostics for evolutionary models of the outer Solar System.

## Acknowledgements
We thank all members of the Deep Ecliptic Survey for their involvement and support. DES observations were carried out at variety of institutions and by many people, as noted explicitly in E05. Much of the data were obtained at NOAO observing facilities, which are operated by AURA, Inc. under a cooperative agreement with the National Science Foundation. A.A.S.G. thanks an anonymous reviewer for comments that improved the manuscript and D. Schleicher for helpful conversations concerning cometary inclination distributions. This work was supported in part by NSF grant AST 07-07609.

**ONLINE ONLY** APPENDIX A

CALCULATION OF THE GEOMETRIC LIKELIHOOD COMPONENT

Solving the integral in equation (15) is computationally time consuming, and numerical integration must be undertaken carefully due to the complicated nature of the functions involved. This Appendix presents the implementation of our calculations for the geometric likelihood component presented in equation (15). For simplicity, the theta dependence notation for latitude and longitude variables is eliminated in this Appendix: for example, $\Delta\beta_{2,k}(\theta_k)$ is represented by $\Delta\beta_{2,k}$.

Integration of equation (15) over the range of latitudes in a given CCD is problematic because the conditional probability density $p(\beta|i)$ can abruptly change slope at the latitudes of the corners of the CCD (and at $\beta = 0°$ for a plane-crossing CCD; see Figure 2 for reference). We define the half height of the $k$th CCD as $H_{1/2,k} = \frac{1}{2} H_k$ and the following variables to represent the latitudes at the corners of the $k$th CCD:

$$b_1 = |\beta_{0,k}| - H_{1/2,k} \qquad b_2 = |\beta_{0,k}| - \Delta\beta_{2,k}$$
$$b_3 = |\beta_{0,k}| + \Delta\beta_{2,k} \qquad b_4 = |\beta_{0,k}| + H_{1/2,k}.$$

(A1)

(For CCDs centered at positive latitudes, $b_1 = \beta_{\min,k}$ and $b_4 = \beta_{\max,k}$.) Let the function $F(x,y)$ represent the following integral, over the latitude range $x$ to $y$:

$$F(x,y) = \int_x^y \Delta\lambda(\beta',\theta_k) p(\beta',i_j) d\beta'.$$

(A2)

The geometric likelihood component for detecting the $j$th object on the $k$th CCD is evaluated by breaking equation (15) into the following cases:

$$\xi_{\text{lat},k,j} = \begin{cases} \sec\theta_k, & i_j = 0° \ \& \ b_1 < i_j \le b_4 \\ c_1[F(b_1, b_2) + F(b_2, b_3) + F(b_3, b_4)], & b_1 > 0 \\ c_1[F(b_1, B_1) + F(B_1, B_2) + F(B_2, B_3) + F(B_3, b_4)], & b_1 \le 0 \le b_4 \end{cases},$$

(A3)

where the normalization factor, $c_1$, is the likelihood of detecting the object over all latitudes,

$$c_1 = \frac{1}{F(-\pi/2, \pi/2)},$$

(A4)

and the integral limits for plane-crossing CCDs vary as a function of imaging location with respect to the reference plane:

$$\left.\begin{matrix} B_1 = 0 \\ B_2 = b_2 \\ B_3 = b_3 \end{matrix}\right\} b_1 < 0 \le b_2, \qquad \left.\begin{matrix} B_1 = b_2 \\ B_2 = 0 \\ B_3 = b_3 \end{matrix}\right\} b_2 < 0 \le b_3, \qquad \left.\begin{matrix} B_1 = b_2 \\ B_2 = b_3 \\ B_3 = 0 \end{matrix}\right\} b_3 < 0 \le b_4.$$

(A5)

The first case in equation (A3) is necessary to account for a divergence of the conditional probability density. The second and third cases represent non-plane crossing CCDs and plane-crossing CCDs, respectively. The limits of integration are adjusted in the third case in order to avoid numerical integration errors at $\beta = 0°$. Representative plots of the resulting geometric likelihood component, and corresponding CCD geometry, are displayed in Figure 2.

Equation (A3) is used in this paper to debias the inclination distribution. However, we have derived an alternative, analytic method for evaluating the geometric likelihood component (the method which was employed in E05). This method takes into account the geometry of each CCD and orbit configuration, assuming no variation in the orbit as a function of the longitude extent of the CCD. It does not require integration, and thus is significantly faster to execute than equation (A3).

The alternative method for deriving the geometric likelihood component considers the following cases: as a function of inclination, object orbits will (i) never reach the latitudes of the observed region, (ii) always be within the latitudes of the observed region, (iii) have both positive and negative latitudes contained within the observed region, (iv) have positive latitudes contained within the observed region and negative latitudes crossing through the observed region, (v) have negative latitudes contained within the observed region and positive latitudes crossing through the observed region, (vi) have both positive and negative latitudes crossing through the observed region, (vii) peak in the latitudes of the observed region, or (viii) cross through the latitudes of the observed region.

For each of these cases, in the order they are listed, the geometric likelihood component can be approximated as

$$\Xi_{lat,k,j} = \begin{cases} 0, & i_j \leq b_1 \\ \sec\theta_k, & |\beta_{0,k}| \leq \Delta\beta_{2,k} \ \& \ i_j \leq \min(|b_2|,|b_3|) \\ \varepsilon_p(\beta_{0,k},i_j)\ \varphi(\beta_{0,k},i_j,\theta_k), & |\beta_{0,k}| \leq H_{1/2,k} \ \& \ i_j \leq \min(|b_1|,b_4) \\ \varepsilon_p(\beta_{0,k},i_j)\ \varphi_+(\beta_{0,k},i_j,\theta_k) + \varepsilon_c(\beta_{0,k},i_j)\ \varphi_-(\beta_{0,k},i_j,\theta_k), & |\beta_{0,k}| \leq H_{1/2,k} \ \& \ |b_1| < i_j \leq b_4 \\ \varepsilon_c(\beta_{0,k},i_j)\ \varphi_+(\beta_{0,k},i_j,\theta_k) + \varepsilon_p(\beta_{0,k},i_j)\ \varphi_-(\beta_{0,k},i_j,\theta_k), & |\beta_{0,k}| \leq H_{1/2,k} \ \& \ b_4 \leq i_j \leq |b_1| \\ \tfrac{1}{2}\varepsilon_c(\beta_{0,k},i_j)\ \varphi(\beta_{0,k},i_j,\theta_k), & |\beta_{0,k}| \leq H_{1/2,k} \ \& \ i_j > \max(|b_1|,b_4) \\ \varepsilon_p(\beta_{0,k},i_j)\ \varphi(\beta_{0,k},i_j,\theta_k), & |\beta_{0,k}| > H_{1/2,k} \ \& \ b_1 < i_j < b_4 \\ \varepsilon_c(\beta_{0,k},i_j), & |\beta_{0,k}| > H_{1/2,k} \ \& \ i_j \geq b_4 \end{cases}$$

(A6)

where $\varepsilon_p(\beta_{0,k}, i_j)$ and $\varepsilon_c(\beta_{0,k}, i_j)$ represent the detection efficiencies for an orbit peaking in or crossing through a CCD, $\varphi(\beta_{0,k}, i_j, \theta_k)$ is the proportion of a tilted CCD relative to a CCD aligned with the plane that does not cross $\beta = 0°$, and $\varphi_+(\beta_{0,k}, i_j, \theta_k)$ and $\varphi_-(\beta_{0,k}, i_j, \theta_k)$ are the proportions above and below the plane for a tilted, plane-crossing CCD relative to an aligned CCD. Explicit definitions for these variables are presented below.

For orbital inclinations that peak within the latitudes of a CCD, the detection efficiencies are

$$\varepsilon_p(\beta_{0,k},i_j) = \begin{cases} 0.5 - \dfrac{1}{\pi}\arcsin\left[\csc i_j \sin|b_1|\right], & |\beta_{0,k}| > H_{1/2,k} \\ 0.5, & |\beta_{0,k}| \leq H_{1/2,k} \end{cases}.$$

(A7)

The maximum peaking efficiency for any CCD that does not span $\beta = 0°$ is 0.5, since only positive or negative latitudes are detectable in these CCDs.

For orbital inclinations that cross through (i) the entire CCD, (ii) either the negative or positive latitudes of a CCD that spans $\beta = 0°$, or (iii) both the negative and positive latitudes of a CCD that spans $\beta = 0°$, the detection efficiencies are

$$\varepsilon_c(\beta_{0,k}, i_j) = \begin{cases} \frac{h}{H_k \pi}\left[\arcsin(\csc i_j \sin b_4) - \arcsin(\csc i_j \sin b_1)\right], & |\beta_{0,k}| > H_{1/2,k} \ \& \ i_j > \max(|b_1|, b_4) \\ \frac{h}{H_k \pi}\arcsin(\csc i_j \sin[\min(|b_1|, b_4)]), & |\beta_{0,k}| \leq H_{1/2,k} \ \& \ i_j \leq \max(|b_1|, b_4) \\ \frac{h}{2H_k \pi}\left[\arcsin(\csc i_j \sin b_4) - \arcsin(\csc i_j \sin b_1)\right], & |\beta_{0,k}| \leq H_{1/2,k} \ \& \ i_j > \max(|b_1|, b_4) \end{cases}$$

(A8)

This method is independent of longitude variations in the CCD; therefore, the factor of $h/H_k$ is used to normalize integrals that extend over a tilted CCD of height $H_k$ to those of a CCD of height $h$ that is aligned with the plane.

We next consider the effect of the angle by which the CCD is tilted with respect to the plane. (See the online-only Appendix C for the effect of tilt angle on detection likelihood.) For a CCD that does not span $\beta = 0°$, $\varphi(\beta_{0,k}, i_j, \theta_k)$ is defined as the proportion of an object's orbit in a tilted CCD relative to an aligned CCD of width $w$ and height $h$:

$$\varphi(\beta_{0,k}, i_j, \theta_k) = \begin{cases} 1, & \theta_k = 0° \\ \frac{h}{w}, & \theta_k = 90° \\ \frac{c_2}{2}(i_j - b_1)^2 \sec\theta_k \csc\theta_k, & b_1 \leq i_j \leq b_2 \\ \frac{c_2}{2}\Delta\beta_{1,k}^2 \sec\theta_k \csc\theta_k + c_2\Delta\lambda_{\max}(i_j - b_2), & b_2 < i_j \leq b_3 \\ c_2 wh - \frac{c_2}{2}(b_4 - i_j)^2 \sec\theta_k \csc\theta_k, & b_3 < i_j \leq b_4 \end{cases}$$

(A9)

where the normalization factor, $c_2$, is

$$c_2 = \frac{1}{w(i_j - b_1)}.$$

(A10)

In equation (A9), the limiting cases of $\theta = 0°$ and $90°$ are necessary to prevent division by zero.

For a CCD spanning $\beta = 0°$, detection of the object at both positive and negative latitudes needs to be considered. Here, we consider the positive and negative latitudes separately and combine them appropriately in equation (A6). We define $\varphi_+(\beta_{0,k}, i_j, \theta_k)$ as the proportion of the positive latitude region in a tilted CCD that spans $\beta = 0°$ in which an object could be detected, relative to that of an aligned CCD:

$$\varphi_+(\beta_{0,k}, i_j, \theta_k) = \begin{cases} 1, & \theta_k = 0° \\ \dfrac{h}{w}, & \theta_k = 90° \\ c_3 \Delta\lambda_{\max} i_j, & |\beta_{0,k}| \leq \Delta\beta_2 \ \& \ i_j \leq b_3 \\ c_3 \Delta\lambda_{\max} b_3 + \dfrac{c_3}{2}\left[\Delta\beta_{1,k}^2 - (b_4 - i_j)^2\right]\csc\theta_k \sec\theta_k, & |\beta_{0,k}| \leq \Delta\beta_2 \ \& \ b_3 < i_j \leq b_4 \\ \dfrac{1}{wb_4}\left(\Delta\lambda_{\max} b_3 + \dfrac{\Delta\beta_{1,k}^2}{2}\csc\theta_k \sec\theta_k\right), & |\beta_{0,k}| \leq \Delta\beta_2 \ \& \ i_j > b_4 \\ \dfrac{c_3}{2}\left[(b_4 + i_j)^2 - |b_1|^2\right]\csc\theta_k \sec\theta_k, & |\beta_{0,k}| > \Delta\beta_2 \ \& \ i_j \leq b_4 \\ c_3 \Delta\lambda_{\max}(i_j - b_2) + \dfrac{c_3}{2}\left[(b_2 + |b_1|)^2 - |b_1|^2\right]\csc\theta_k \sec\theta_k, & |\beta_{0,k}| > \Delta\beta_2 \ \& \ |b_2| < i_j \leq b_3 \\ c_3 wh - \dfrac{c_3}{2}\left[|b_1|^2 + (b_4 - i_j)^2\right]\csc\theta_k \sec\theta_k. & |\beta_{0,k}| > \Delta\beta_2 \ \& \ b_3 < i_j \leq b_4 \\ \dfrac{1}{wb_4}\left(wh - \dfrac{|b_1|^2}{2}\csc\theta_k \sec\theta_k\right), & |\beta_{0,k}| > \Delta\beta_2 \ \& \ i_j > b_4 \end{cases}$$

(A11)

where the normalization factor, $c_3$, is

$$c_3 = \begin{cases} \dfrac{1}{w}, & i_j = 0° \\ \dfrac{1}{wi_j}, & i_j \neq 0° \end{cases}.$$

(A12)

We similarly define $\varphi_-(\beta_{0,k}, i_j, \theta_k)$ as the proportion of the negative latitude region in a tilted CCD that spans $\beta = 0°$ in which an object could be detected, relative to that of a CCD aligned with the plane:

$$\varphi_-(\beta_{0,k}, i_j, \theta_k) = \begin{cases} 1, & \theta_k = 0° \\ \dfrac{h}{w}, & \theta_k = 90° \\ \dfrac{c_3}{2}\left(2i|b_1| - i_j^2\right)\csc\theta_k \sec\theta_k, & |\beta_{0,k}| > \Delta\beta_2 \ \& \ i_j \leq |b_1| \\ \dfrac{|b_1|}{2w}\csc\theta_k \sec\theta_k, & |\beta_{0,k}| > \Delta\beta_2 \ \& \ i_j > |b_1| \\ \dfrac{1}{wb_4}\left(\Delta\lambda_{\max} b_3 + \dfrac{\Delta\beta_{1,k}^2}{2}\csc\theta_k \sec\theta_k\right), & |\beta_{0,k}| \leq \Delta\beta_2 \ \& \ i_j > b_4 \\ c_3 \Delta\lambda_{\max} i_j, & |\beta_{0,k}| \leq \Delta\beta_2 \ \& \ i_j \leq |b_2| \\ c_3 \Delta\lambda_{\max}|b_2| + \dfrac{c_3}{2}\left[(|b_1|-|b_2|)^2 - (|b_1|-i)^2\right]\csc\theta_k \sec\theta_k, & |\beta_{0,k}| \leq \Delta\beta_2 \ \& \ |b_2| < i_j \leq |b_1| \\ \dfrac{1}{wb_1}\left(\Delta\lambda_{\max}|b_2| + \dfrac{1}{2}\left[(|b_1|-|b_2|)^2\right]\csc\theta_k \sec\theta_k\right), & |\beta_{0,k}| \leq \Delta\beta_2 \ \& \ i_j > |b_1| \end{cases}.$$

(A13)

The alternative method for calculation of the geometric likelihood component given by equation (A6) is faster than evaluation of the integral form in equation (A3) by a factor of a few hundred, running *Mathematica 5.2* on an Apple PowerBook G4 with a 1.5 GHz processor. This alternative method is accurate for CCDs aligned with the plane ($\theta = 0°$, 90°). For tilted CCDs, however, the assumption of no variation in an object's orbit as a function of the longitude extent of the CCD leads to discrepancies. The discrepancy between equations (B3) and (B6) increases with tilt angle, peaking at $\theta = \arctan(h/w) = 63.44°$. The discrepancy is most severe for plane-crossing CCDs, since in these cases the likelihood varies steeply as a function of orbital inclination (c.f. Figure 2).

For CCDs in DES search frames, the tilt angles range from 0.05° to 23.56°, with a latitude distribution plotted in online-only Figure 12. The maximum discrepancy in the likelihood component between equations (A3) and (A6) is 0.06 – a 25% variation from the likelihood itself. The average percent different between methods over all DES CCDs is 3.8%. Thus, the likelihood factors for inclination debiasing employed in E05 vary only slightly from the more rigorous, slower method used in this work.

## APPENDIX B

## OBJECT INCLINATION WITH RESPECT TO THE KBP

The following equation is used to convert between the ecliptic inclination of an object, $i_e$, and the object's inclination with respect to a different plane, $i_{plane}$:

$$i_{plane} = \frac{\pi}{2} - \sin^{-1}\left[\sin\left(\frac{\pi}{2} - i_e\right)\cos I - \sin I \cos\left(\frac{\pi}{2} - i_e\right)\sin\left(\frac{3\pi}{2} + \Omega_{object} - \Omega_{plane}\right)\right],$$

(B1)

where $\Omega_{object}$ is the longitude of the ascending node of the object, $I$ is the inclination of the new plane, and $\Omega_{plane}$ is the longitude of the ascending node of the new plane, all with respect to the ecliptic. For inclinations with respect to the KBP, $i_K$, we use the solution from E05 of $I = 1.74°$ and $\Omega_{plane} = 99.2°$.

## **ONLINE ONLY** APPENDIX C

## EFFECT OF CCD TILT ANGLE ON DETECTION LIKELIHOOD

To investigate the importance of the angles at which survey CCDs are tilted with respect to the reference plane, we calculate detection likelihoods (equation 20) for all DES objects assuming $\theta = 0°$. The resulting likelihoods differ by < 1.5% from the values obtained when accounting for tilt angle (online-only Figure 13). The difference is largest for objects having low inclinations. Overall, the effect of the CCD tilt angles can be considered minimal for the inclination distribution analyses presented here. Note that our method considers relative likelihoods of detection, and the tilt angles may play a more important role for future calculations of absolute detection likelihood.

TABLE 1

SAMPLE STATISTICS

| Sample[a] | No. objects[b] | No. objects in Elliot et al. (2005)[c] |
|---|---|---|
| All DES | 482 | 373 |
|   Inclination error $\sigma_i \leq 0.5°$ | 361 | 245 |
|     Centaurs | 17 | 5 |
|     DES KBOs | 344 | 240 |
|       Classified[d] | 295 | 174 |
|         Resonant KBOs[e] | 104 | 52 |
|           3:2 $e$ | 51 | 29 |
|           Other Resonant | 53 | 23 |
|             7:4 $e^3$ | 12 | 7 |
|         Nonresonant KBOs | 191 | 122 |
|           Classical | 150 | 92 |
|           Scattered | 41 | 30 |
|             Scattered Near | 32 | 21 |
|             Scattered Extended | 9 | 9 |
|       Unclassified | 49 | 66 |

[a] Objects designated by the MPC that were discovered on valid DES search fields; indentation denotes subsamples.

[b] Values as of 2007 October 24, which are used for analyses in this work.

[c] Values as of 2003 December 31 that were used in E05 and are provided here for reference. These samples considered an additional criterion of KBOs with heliocentric distance < 30 AU, which eliminated seven objects from the sample of objects with low inclination error.

[d] We consider objects with classification quality $\geq 2$, following the classification scheme presented in E05. Only 7 objects have quality 2 and the remainder are quality 3.

[e] The two Resonant subsamples with the greatest number of objects are listed as distinct samples. There are six objects or fewer in each of 18 other occupied Resonant subsamples.

# TABLE 2
## FITTING RESULTS FOR UNBIASED INCLINATION DISTRIBUTIONS

| Sample[a] | Functions and best-fit values[b] | | |
|---|---|---|---|
| | sin $i$ multiplied by single Gaussian ($f_G(i)$; Eq. 29) | | |
| | $\mu_1$ | $\sigma_1$ | $V\sqrt{N}$ (rejection %)[c] |
| 3:2 $e$ | 0 | $10.7^{+2.0}_{-2.3}$ | 1.29 (64.4) |
| 7:4 | 0 | $2.4^{+2.2}_{-1.1}$ | 0.98 (19.7) |
| Other Resonant | 0 | $11.0^{+2.8}_{-2.5}$ | 1.12 (39.6) |
| Scattered Extended | 0 | $13.2^{+9.1}_{-5.0}$ | 1.27 (62.0) |
| 3:2 $e$[d] | $6.5^{+3.8}_{-6.5}$ | $8.5^{+1.8}_{-2.0}$ | 1.26 (59.8) |
| 7:4 | $2.3^{+2.1}_{-2.3}$ | $1.5^{+2.2}_{-1.0}$ | 0.93 (13.7) |
| Other Resonant | $6.0^{+3.8}_{-6.0}$ | $8.7^{+1.8}_{-2.0}$ | 1.26 (59.7) |
| Scattered[d] | $19.1^{+3.9}_{-3.6}$ | $6.9^{+4.1}_{-2.7}$ | 1.06 (30.9) |
| Scattered Near[d] | $20.3^{+4.5}_{-4.0}$ | $6.8^{+4.8}_{-3.0}$ | 1.04 (27.1) |
| Scattered Extended[d] | $17.2^{+4.2}_{-3.3}$ | $2.9^{+10.6}_{-2.3}$ | 0.94 (15.1) |
| | sin $i$ multiplied by generalized Lorentzian ($f_L(i)$; Eq. 30) | | |
| | $I_1$ | $g_1$[e] | $V\sqrt{N}$ (rejection %)[c] |
| 3:2 $e$ | $8.8^{+1.2}_{-1.7}$ | $6.4^{+3.4}_{-1.5}$ | 1.27 (61.3) |
| 7:4[d] | $2.2^{+1.4}_{-0.8}$ | $20.0^{+>100}_{-16.8}$ | 0.88 (8.6) |
| Other Resonant[d] | $7.0^{+1.9}_{-1.7}$ | $4.4^{+2.1}_{-1.0}$ | 1.06 (30.6) |
| Scattered | $14.1^{+2.3}_{-1.4}$ | $12.1^{+>100}_{-5.5}$ | 1.23 (57.4) |
| Scattered Near | $15.0^{+2.7}_{-1.8}$ | $18.7^{+>100}_{-12.2}$ | 1.05 (28.6) |
| Scattered Extended | $10.9^{+6.6}_{-2.5}$ | $21.0^{+>100}_{-17.9}$ | 1.21 (52.4) |
| | sin $i$ multiplied by sum of two Gaussians ($f_{2G}(i)$; Eq. 31) | | |
| | $a_1$ | $\sigma_2$ | $\sigma_3$ | $V\sqrt{N}$ (rejection %)[c] |
| DES KBOs | $0.18 \pm 0.06$ | $1.8^{+0.8}_{-0.6}$ | $13.6 \pm 0.9$ | 1.32 (67.9) |
| Classical | $0.53^{+0.17}_{-0.15}$ | $2.0^{+0.6}_{-0.5}$ | $8.1^{+2.6}_{-2.1}$ | 0.99 (20.6) |
| Unclassified | $0.51 \pm 0.24$ | $1.3^{+0.9}_{-0.6}$ | $4.8^{+3.6}_{-1.8}$ | 1.02 (25.4) |
| | sin $i$ multiplied by a Gaussian plus Lorentzian ($f_{GL}(i)$; Eq. 32) | | | |
| | $a_2$ | $\sigma_4$ | $I_2$ | $g_2$[e] | $V\sqrt{N}$ (rejection %)[c] |
| DES KBOs[d] | $0.25 \pm 0.07$ | $2.3^{+0.9}_{-0.7}$ | $11.2 \pm 0.9$ | $7.0^{+2.5}_{-1.4}$ | 1.14 (43.0) |
| Classical[d] | $0.58 \pm 0.15$ | $1.9^{+0.6}_{-0.4}$ | $6.6^{+1.8}_{-1.4}$ | $9.2^{+>100}_{-4.7}$ | 0.96 (17.4) |
| Unclassified | $0.57^{+0.22}_{-0.24}$ | $1.4^{+0.8}_{-0.6}$ | $3.9^{+3.3}_{-1.5}$ | $5.9^{+>100}_{-3.4}$ | 1.03 (26.9) |

[a] Samples correspond to those listed in Table 1.
[b] The best-fit values, ±1 sigma, from our Monte Carlo method applied to each sample and each function.
[c] The statistic from Kuiper's variant of the K-S test for the best-fit parameters, and the corresponding confidence level at which we can reject the hypothesis that the sample and the fit are drawn from the same intrinsic distribution.
[d] Fits with the lowest confidence level of rejection for each sample. Other fits listed in the table are acceptable at the 1-sigma level and cannot be ruled out.
[e] Values for the Lorentzian power parameter, $g$, were only measured up to the best-fit value plus 100. See Figure 9 for plots that provide an example of the large, 1-sigma, best-fit parameter space for $g$.

TABLE 3

REJECTED FITS FOR A GAUSSIAN MULTIPLIED BY SIN $i$

| Sample[a] | $\sigma_i$ (deg)[b] | $V\sqrt{N}$ [c] | Confidence level of rejection (%)[c] |
|---|---|---|---|
| DES KBOs | 13.1 | 2.77 | 97.4 |
| Classical | 2.6 | 2.90 | 98.8 |
| Scattered | 16.4 | 1.73 | 95.3 |
| Scattered Near | 17.7 | 1.70 | 94.6 |
| Unclassified | 2.4 | 1.82 | 97.3 |

[a] Samples correspond to those listed in Table 1.
[b] The best-fit values from our Monte Carlo method using equation (29) with $\mu_i=0$.
[c] The statistic from Kuiper's variant of the K-S test for the best-fit value, and the corresponding confidence level at which we can reject the hypothesis that the two samples were drawn from the same intrinsic distribution. Percentages > 84.1% and > 97.7% are beyond the 1- and 2-sigma levels respectively.

TABLE 4

COMPARISON BETWEEN KBO, CENTAUR, AND COMET INCLINATIONS

| Sample[a] | Comparison sample[a] | | | |
|---|---|---|---|---|
| | Centaurs | | JFCs | |
| | $V\sqrt{N}$ [b] | Confidence level of rejection (%)[b] | $V\sqrt{N}$ [b] | Confidence level of rejection (%)[b] |
| DES KBOs | 1.39 | 80.5 | 3.15 | >99.9 |
| 3:2 $e$ | 2.15 | 99.9 | 2.34 | >99.9 |
| Other Resonant | 1.98 | 99.6 | 2.08 | 99.7 |
| Classical | 2.93 | >99.9 | 6.18 | >99.9 |
| Scattered | 1.84 | 99.1 | 2.15 | 99.9 |
| Scattered Near | 2.01 | 99.7 | 2.31 | >99.9 |
| Scattered Extended | 1.31 | 79.9 | 1.55 | 92.7 |
| JFCs | 1.68 | 95.8 | – | – |

[a] Samples correspond to those listed in Table 1. Jupiter-family comets (JFCs) are the 256 comets designated by the MPC as of 2008 October 30 with $2 < T_J < 3.05$ and $q < 2.5$ AU. The comet sample we use has been integrated to a common, arbitrary epoch (2003 09 18 0:00) and serves as a snapshot for the inclination distribution.

[b] The statistic from Kuiper's variant of the K-S test, and the corresponding confidence level at which we can reject the null hypothesis that the two samples were drawn from the same intrinsic distribution. Percentages > 84.1%, 97.7%, and 99.9% are beyond the 1-, 2-, and 3-sigma levels respectively.

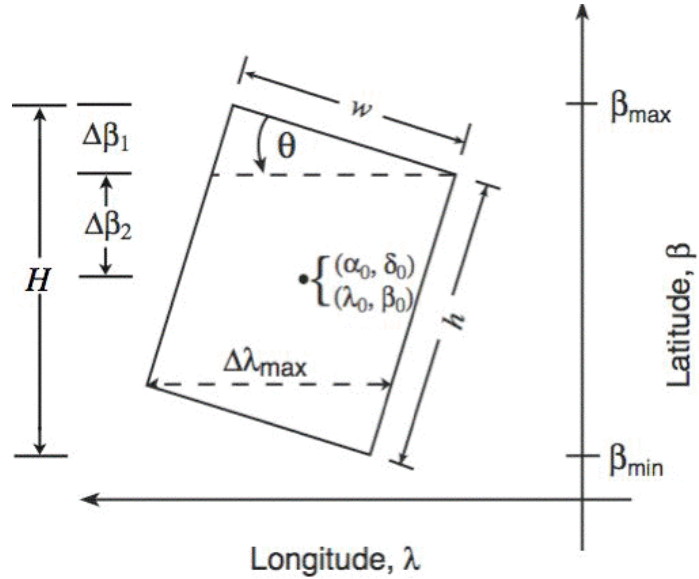

Figure 1. Schematic of a CCD, eight of which comprise one DES Mosaic search frame. Each CCD is a rectangular element of solid angle, having angular height and width of $h$ and $w$ respectively. The CCDs are aligned with the right ascension and declination $(\alpha, \delta)$ coordinate axes. We consider the transformation of this solid angle to another spherical coordinate system that has longitude and latitude coordinates $(\lambda, \beta)$. The center of the CCD is thus transformed from coordinates $(\alpha_0, \delta_0)$ to $(\lambda_0, \beta_0)$. The CCD is assumed to be far from the pole of either coordinate system, so that the coordinate transformation is approximated by a rotation of the rectangular element by angle $\theta$, defined as the position angle (measured from North through East) of the $\beta$ axis in the $(\alpha, \delta)$ coordinate system. The full angular extent of the CCD in latitude is denoted by $H$, with minimum and maximum values $\beta_{min}$ and $\beta_{max}$. The dashed lines are of constant latitude, and these break the CCD into subelements over which it is convenient to carry out integrations. Useful measures $\Delta\lambda_{max}$, $\Delta\beta_1$, and $\Delta\beta_2$ are also labeled. (Figure adapted from E05.)

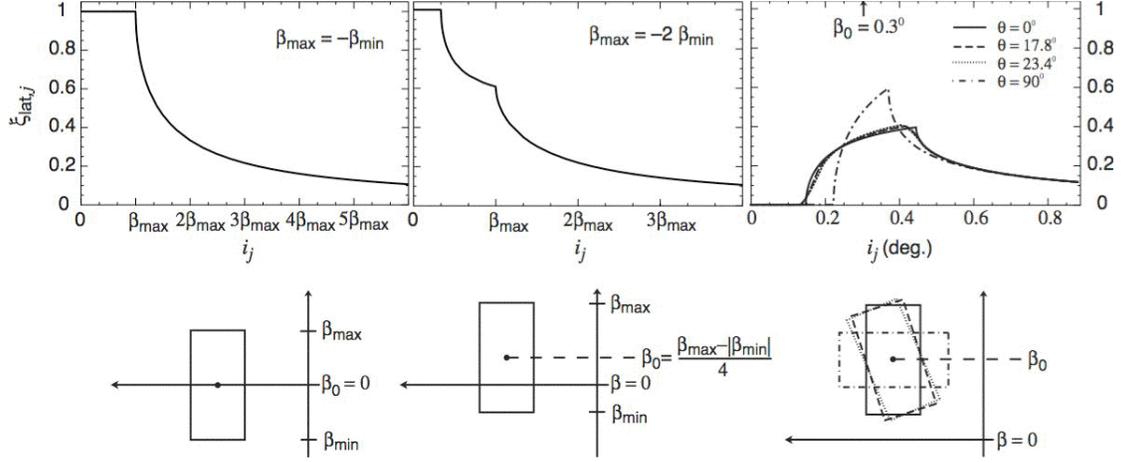

Figure 2. Plots of the normalized, geometric likelihood component, $\xi_{\text{lat},j}$, for detecting an object having orbital inclination $i_j$ in a CCD of various configurations. Schematic diagrams of the CCD are displayed below the corresponding likelihood plot. The likelihood component is normalized by solving equation (15) for a DES CCD integrated over all latitudes. (*left*) A plane-crossing CCD that has equal areas of positive and negative latitudes, $|\beta_{\min}| = |\beta_{\max}|$. In this case, $\xi_{\text{lat},j}$ is 1 when the orbital inclination is fully contained within the frame. The likelihood decreases with increasing inclination, corresponding to the orbit crossing through the CCD latitudes and having an increasingly larger fraction of the orbit outside of the CCD. (*middle*) A plane-crossing CCD with unequal areas of positive and negative latitude. To illustrate, we have chosen the specific case when the center of the CCD is a quarter of the total CCD height above $\beta = 0°$. The normalized likelihood is 1 only when the orbital inclination is contained within both positive and negative latitudes of the CCD. Then, there are two transitions: (i) when the inclination has increased beyond one, but not both, of the CCD latitude limits, and (ii) when the orbital inclination is greater than the CCD latitude farthest from $\beta = 0°$ (the maximum of $|\beta_{\min}|$ or $|\beta_{\max}|$), after which point the likelihood decreases accordingly. (*right*) A non-plane-crossing CCD, with examples of CCD tilts $\theta = 0°$, 17.8°, 23.4°, and 90°. The intermediate tilt angles are chosen to illustrate the mean and maximum of DES frames, 17.8° and 23.4° respectively. Because $\beta_{\min}$ and $\beta_{\max}$ vary as functions of $\theta$, we set the CCD to the size of a DES Mosaic CCD, $w = 0.148°$ and $h = 0.296°$. For this example, we assume a latitude center for the CCD of $\beta_0 = 0.3°$. This selection results in positive latitudes: the cases for the corresponding negative latitudes are identical. There is no likelihood of detection until the orbital inclination reaches the CCD latitude closest to $\beta = 0°$ (the minimum of $|\beta_{\min}|$ or $|\beta_{\max}|$). The likelihood increases as the orbit peaks within the CCD latitudes. Once the inclination is greater than the CCD latitude farthest from $\beta = 0°$, the likelihood decreases accordingly. At $\theta = 0°$ the extent of the CCD in latitude is equal to $h$, while at $\theta = 90°$ it is equal to $w$.

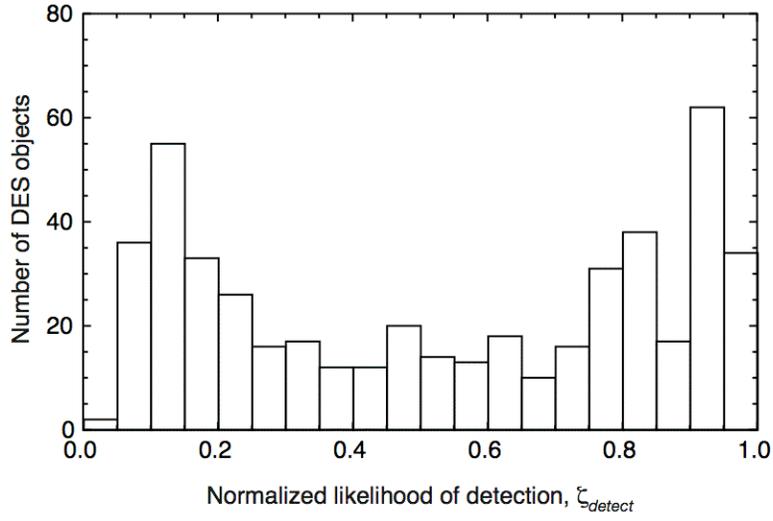

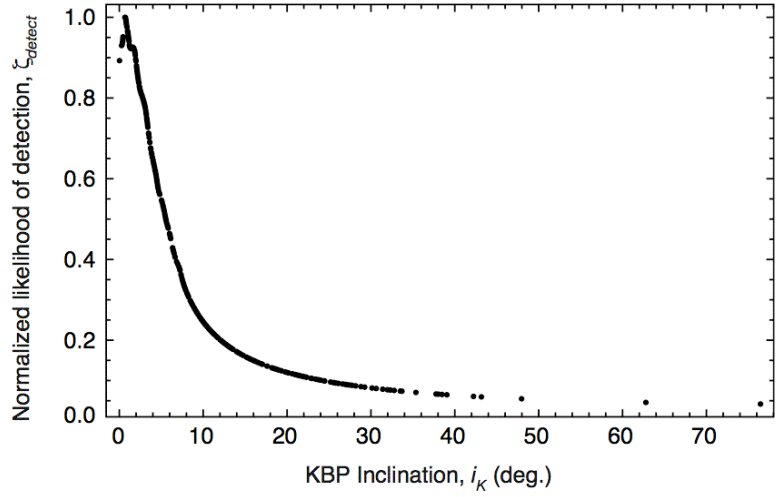

Figure 3. The relative likelihood of detecting each of the 482 KBOs and Centaurs discovered on valid DES search frames. (a) The number of objects as a function of detection likelihood, $\zeta_{detect}$ (equation 20). We consider relative likelihoods; therefore, the values are normalized to the maximum likelihood obtained for any individual object. (b) The normalized likelihood of detection as a function of object inclination with respect to the KBP.

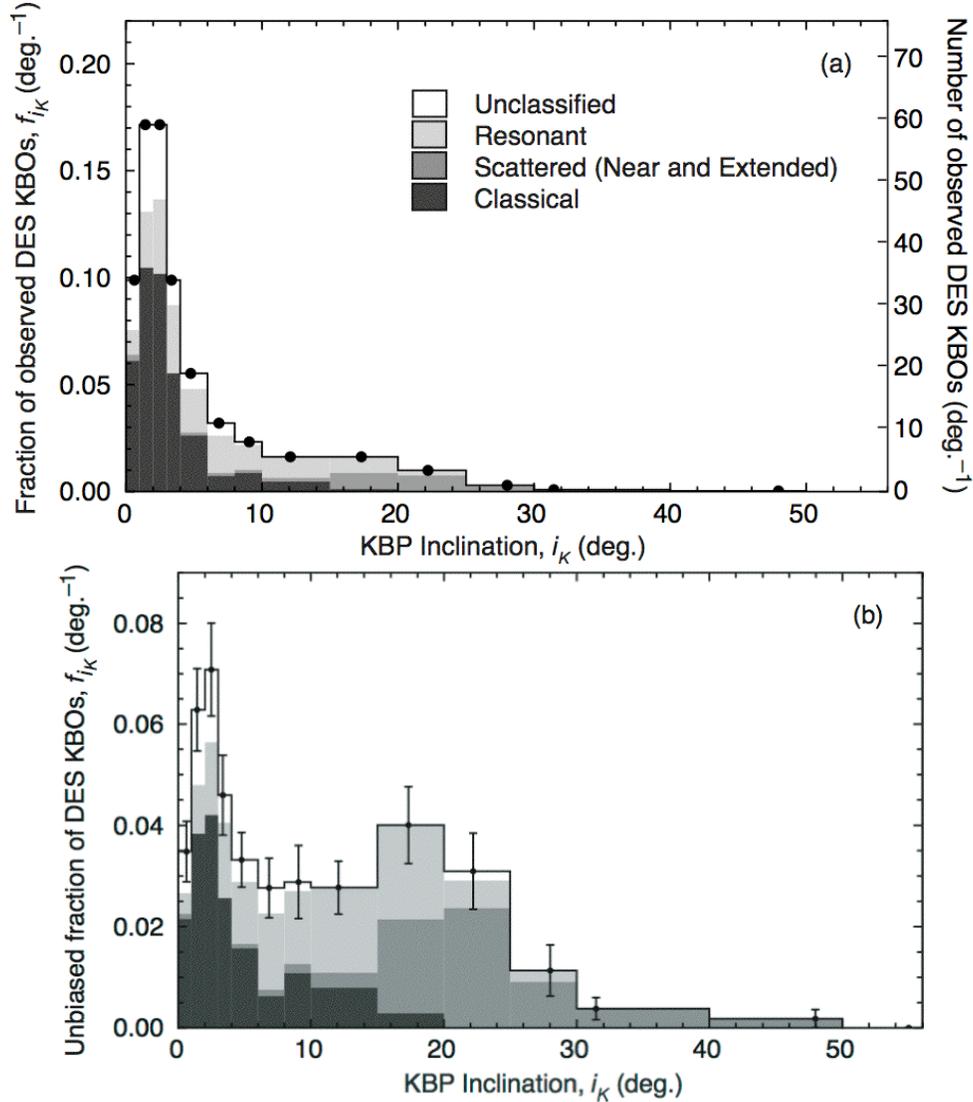

Figure 4. Inclination distributions for DES KBOs having inclination error ≤ 0.5° as of 2007 October 24 (344 objects). (*a*) The observed fraction of objects (left axis) and number of objects (right axis) per degree of inclination with respect to the KBP. Bin sizes are 1° from 0° to 4°, 2° from 4° to 10°, 5° from 10° to 25°, and one 10° bin from 25° to 55° (following Millis *et al.* 2002; E05, for bins < 25°). Data points are plotted at the average inclination for the objects in each bin. The data points are enlarged to be more clearly visible for higher-inclination bins in which few objects were detected. Each bin is shaded to reflect the proportion of KBOs by classification: Unclassified objects are represented by white areas, Resonant objects are represented by light gray, Scattered (Near and Extended) objects are represented by medium gray, and Classical objects are represented by dark gray. (*b*) The unbiased fraction of objects per degree of inclination with respect to the KBP. Data points and error bars are determined following the debiasing procedure presented in §2. The debiased plot represents only the observed DES KBOs and does not consider bias effects that may have caused objects to not be detected; therefore, this plot should not be considered as a representation of the relative populations of different KBO classes.

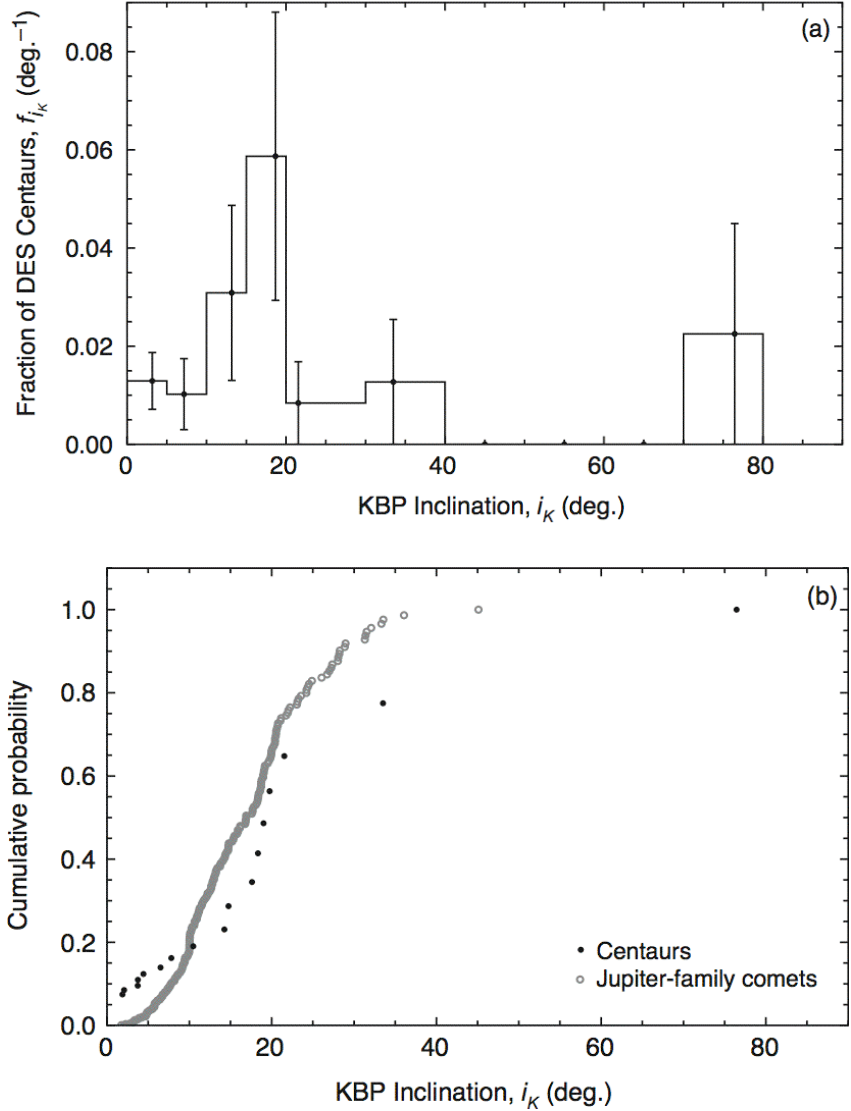

Figure 5. Unbiased inclination distribution for all DES Centaurs having inclination error ≤ 0.5° as of 2007 October 24 (17 objects). (*a*) Data points and error bars represent the fraction of objects per degree of inclination with respect to the KBP, following the debiasing procedure presented in §2. Due to the paucity of objects, bin sizes are 5° from 0° to 20°, and 20° from 20° to 80°. Data points are plotted at the average inclination for the objects in each bin. (*b*) The cumulative distribution of Centaurs as a function of inclination with respect to the KBP (solid black dots). From 0 to 1, this plot represents the fraction of Centaurs having inclinations at or below the corresponding abscissa value. The steep increase in objects near 20° is apparent. For comparison, the cumulative distribution of JFCs is also shown (open gray circles). The probability that the current inclination distributions of Centaurs and JFCs are derived from the same parent distribution cannot be rejected at the 2-sigma level (see Table 4).

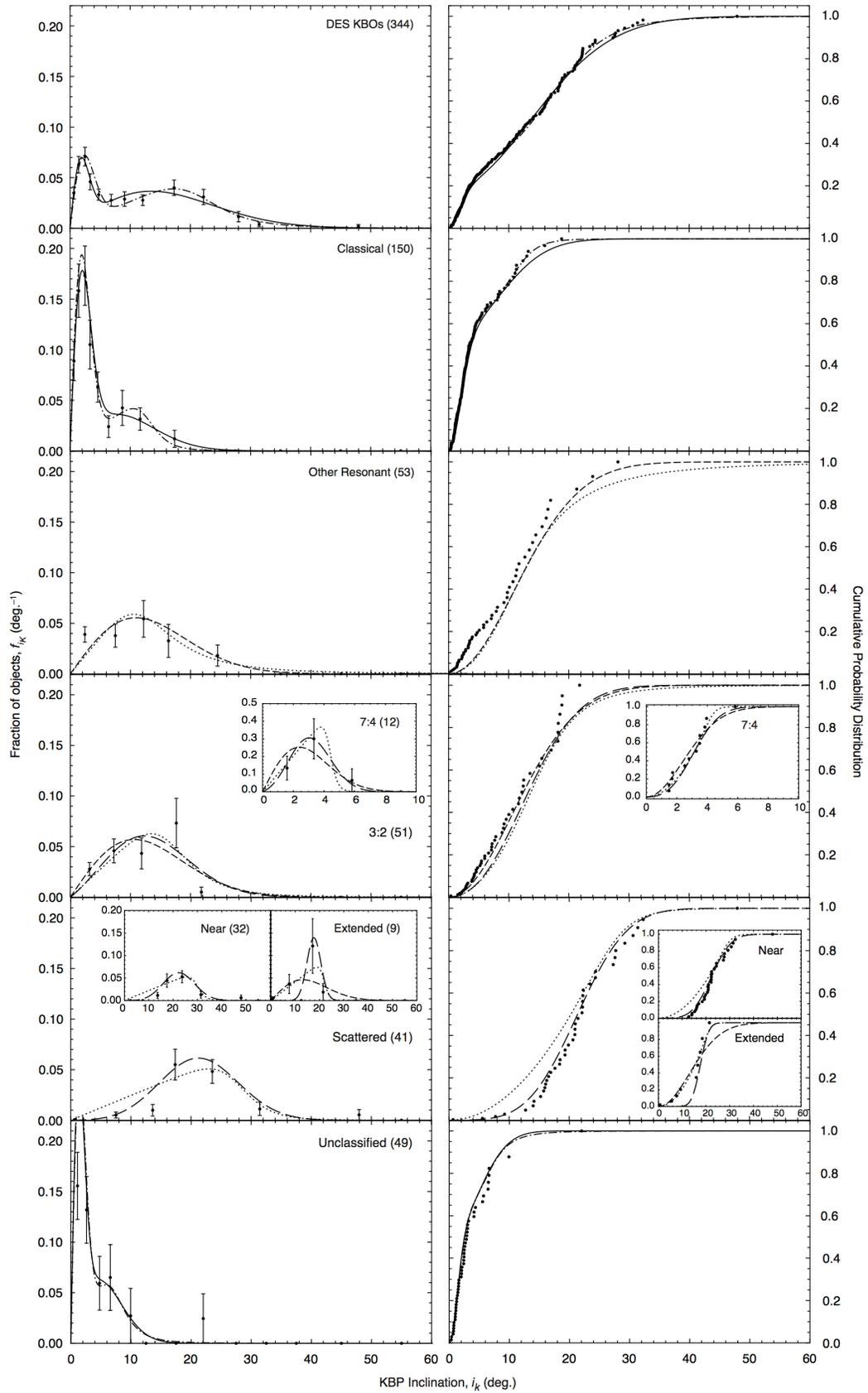

Figure 6. Unbiased inclination distributions for selected samples from Table 1. On the left, the fraction of the population per degree of inclination with respect to the KBP is plotted as data points with error bars for each sample. The sample names are provided in the plots, with the number of objects in the sample in parentheses. Bin sizes vary, in order to ensure at least a few objects in each bin. For DES KBOs and Classical objects, bin sizes are 1º from 0º to 4º, 2º from 4º to 10º, 5º from 10º to 30º, and 10º bins from 30º to 80º; for 3:2, Other Resonant, Scattered Near, Scattered Extended, and Scattered, bin sizes are 5º from 0º to 20º and 10º from 20º to 80º; for 7:4 and Unclassified, bin sizes are 2º from 0º to 10º, 5º from 10º to 40º, and 10º from 40º to 80º. The cumulative distributions (unbinned) for each sample are shown on the right. The best-fit functional forms for the inclination distribution, from Table 2, are overplotted on the data. Short-dashed lines represent sin $i$ multiplied by a single Gaussian (equation 29 with $\mu = 0$), long-dashed lines are sin $i$ multiplied by a single Gaussian offset from the plane (equation 29 with $\mu \neq 0$), dotted lines are sin $i$ multiplied by a generalized Lorentzian (equation 30), solid lines are sin $i$ multiplied by the sum of two Gaussians (equation 31), and dot-dashed lines are sin $i$ multiplied by a Gaussian plus generalized Lorentzian (equation 32).

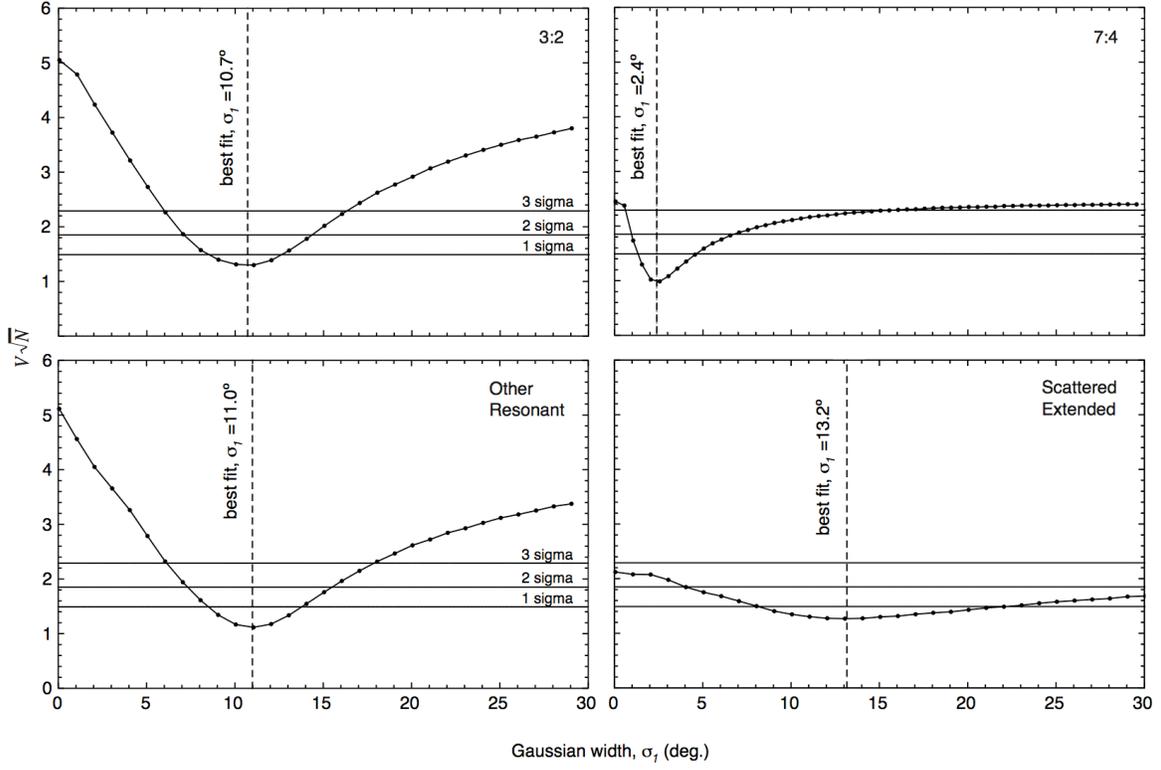

Figure. 7. The $V\sqrt{N}$ statistic as a function of $\sigma_1$ for $\sin i$ multiplied by a single Gaussian centered on the plane (equation 29 with $\mu_1=0$) for KBO samples with acceptable fits. The solid lines denote the calibrated confidence levels at which we can reject the hypothesis that the model and the sample are drawn from the same distribution. The best-fit values are denoted by dashed lines, and our error bars (listed in Table 2) encompass the range for which $V\sqrt{N}$ is less than one sigma.

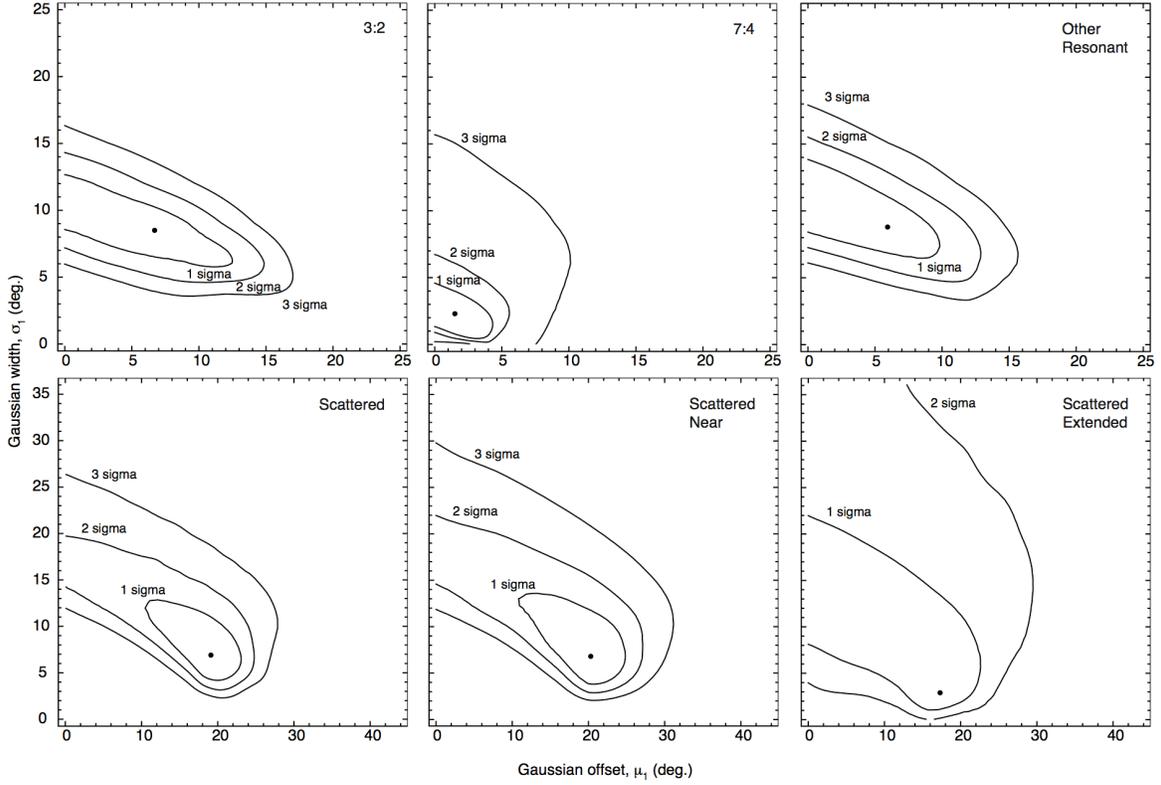

Figure 8. The $V\sqrt{N}$ statistic as a function of $\mu_1$ and $\sigma_1$ for sin $i$ multiplied by a single Gaussian (equation 29 with $\mu_1 \neq 0$) for KBO samples with acceptable fits. The best-fit values from Table 2 are shown as dots and the solid contour lines denote the confidence levels. One-sigma error bars are determined for the best-fit values by calculating $V\sqrt{N}$ for each parameter while keeping the other fixed, similar to Figures 7, 9b, and 9c.

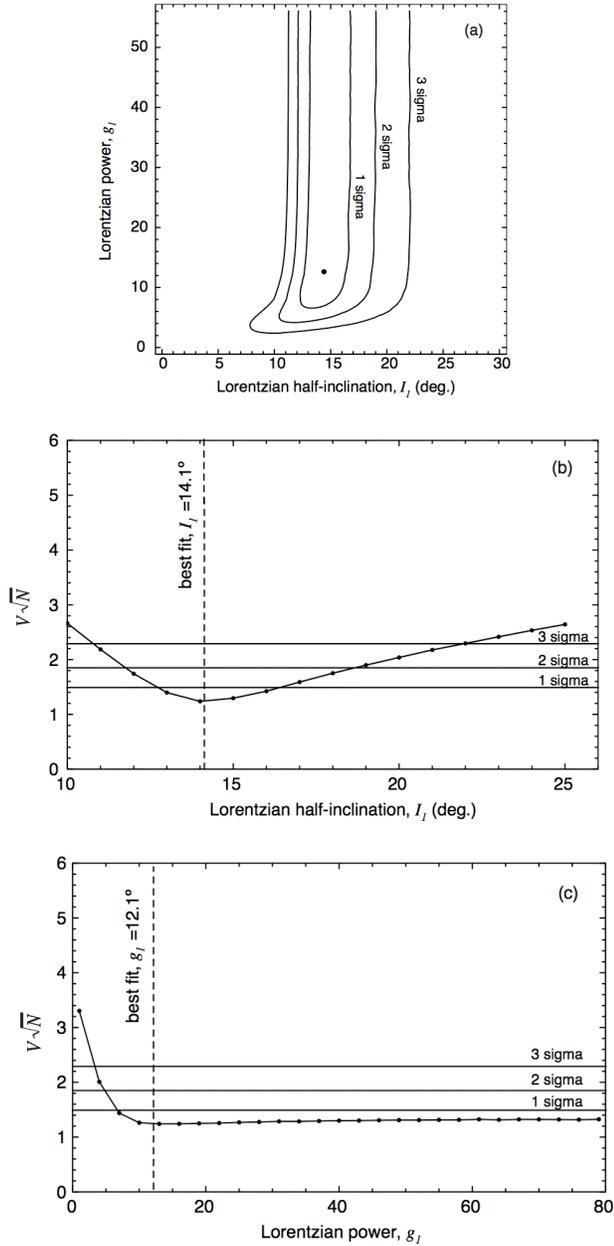

Figure 9. (*a*) The $V\sqrt{N}$ statistic as a function of $I_1$ and $g_1$ for $\sin i$ multiplied by a Lorentzian (equation 30) for the sample of Scattered KBOs. The solid contour lines denote the confidence levels of the fits. The dot represents the best-fit value listed in Table 2, for which the hypothesis that the model and the sample were drawn from the same intrinsic distribution is rejected at a confidence level of only 57.4%. This plot illustrates the large range in $g_1$ over which fits are acceptable. (*b*) The $V\sqrt{N}$ statistic for $I_1$ while holding $g_1$ at the best-fit value. (*c*) The $V\sqrt{N}$ statistic for $g_1$ while holding $I_1$ at the best-fit value.

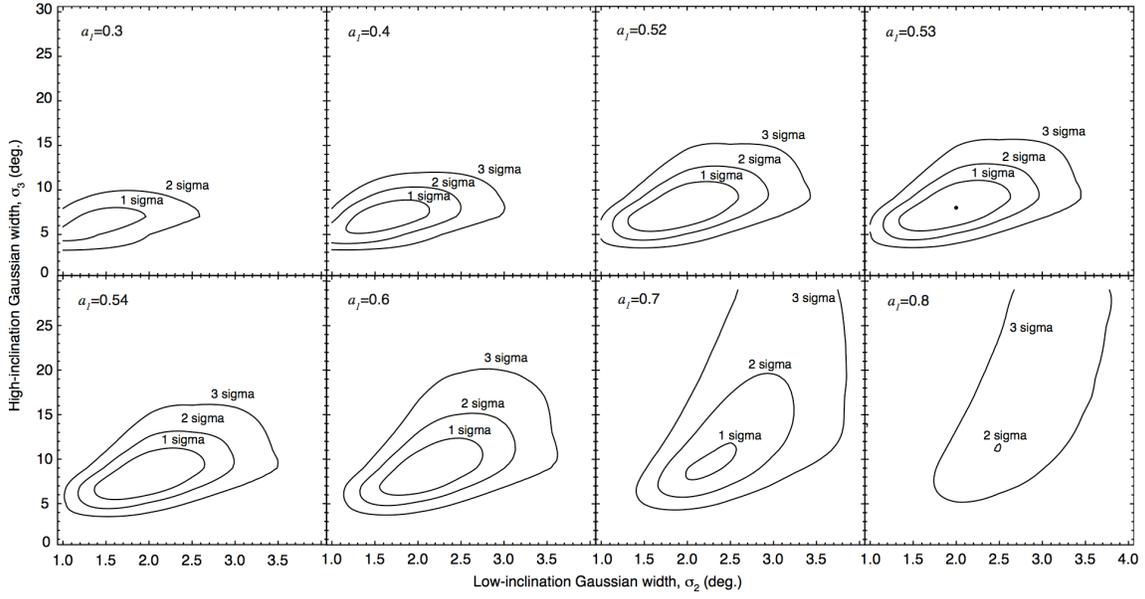

Figure 10. The $V\sqrt{N}$ statistic as a function of $a_1$, $\sigma_2$, and $\sigma_3$ for the sum of two Gaussians multiplied by $\sin i$ (equation 31) for the sample of Classical KBOs. The solid contour lines denote the confidence levels of the fit, and the best-fit value from Table 2 is represented by a dot. For the best fit, the hypothesis that the model and the sample were drawn from the same intrinsic distribution is rejected at a confidence level of only 20.6%.

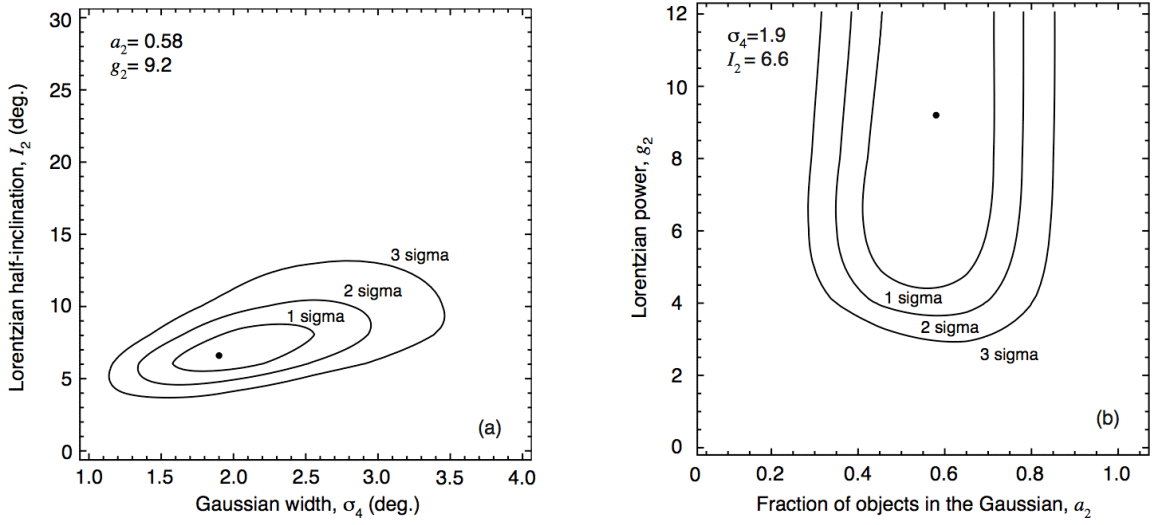

Figure 11. The $V\sqrt{N}$ statistic for a Gaussian plus a Lorentzian multiplied by sin $i$ (equation 32) for the sample of Classical KBOs. The solid contour lines denote the confidence levels of the fit, and the best-fit value from Table 2 is represented by a dot. Because this is a four-parameter fit, we provide two plots to represent the results: (*a*) contours for $I_2$ versus $\sigma_4$ given the best-fit values of $g_2$ and $a_2$ (this plot is comparable to the best fit in Figure 10), and (*b*) contours for $g_2$ versus $a_2$ given the best-fit values of $\sigma_4$ and $I_2$. For the best fit, the hypothesis that the model and the sample were drawn from the same intrinsic distribution is rejected at a confidence level of only 17.4%.

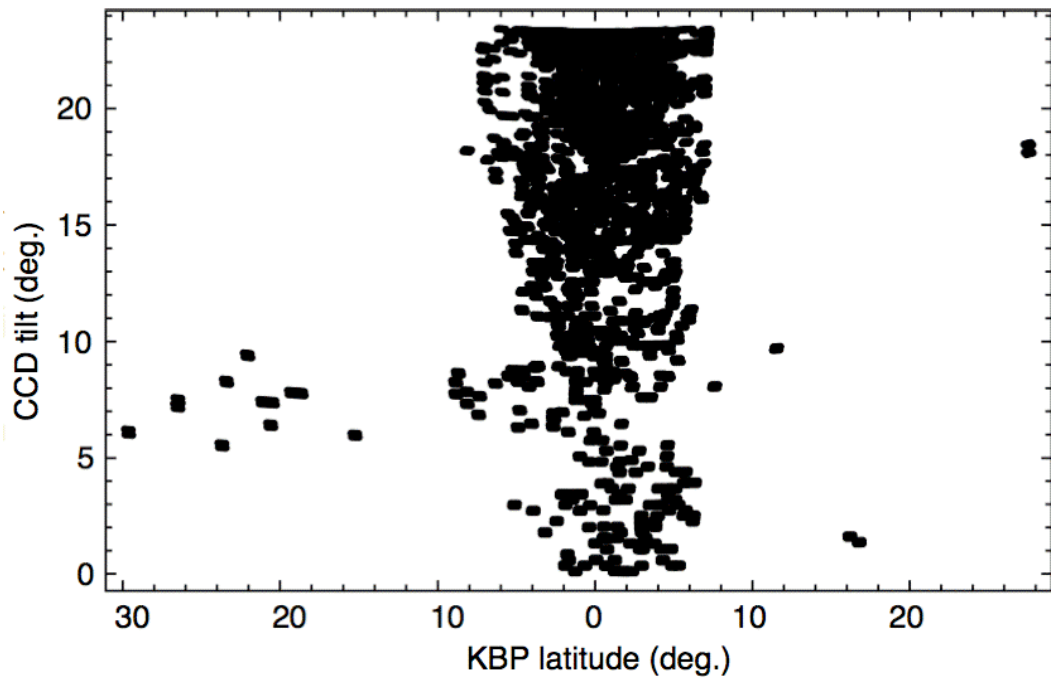

**ONLINE ONLY** Figure 12. Tilt angle, $\theta$, with respect to the KBP latitude for all CCDs on valid DES search frames. There are 19056 CCDs in valid frames, each represented here by a black dot. The median tilt angle is 19.55°. Considering tilt angles with respect to the ecliptic changes this plot only slightly – in that case, the maximum and median tilts are 23.44° and 19.3° and the maximum ecliptic latitude is 29.07°.

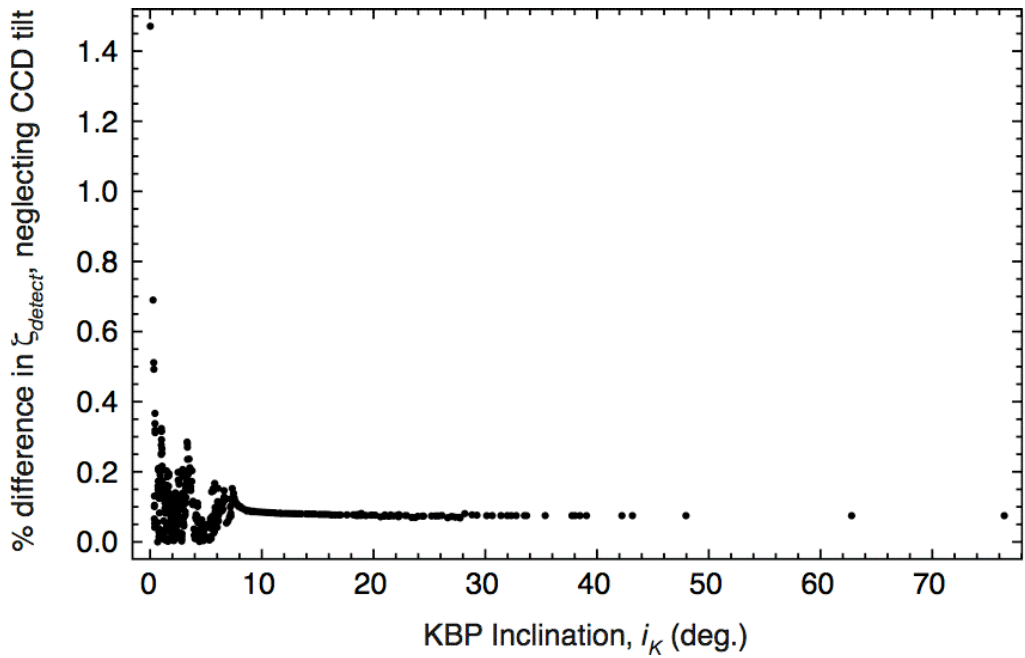

ONLINE ONLY Figure 13. The percent difference in detection likelihood, $\zeta_{detect}$, between accounting for the tilt angles of DES search frames and assuming tilt angle = $0°$. Values are plotted for each of the 482 DES KBOs and Centaurs discovered on valid search frames as a function of inclination with respect to the KBP.